\documentclass[preprint,journal]{vgtc}       % preprint (journal style)

\ieeedoi{10.1109/TVCG.2019.2934335}

\pdfoutput=1\relax                   % create PDFs from pdfLaTeX
\usepackage{graphicx}                % allow us to embed graphics files
\DeclareGraphicsExtensions{.pdf,.png,.jpg,.jpeg} % for pdflatex we expect .pdf, .png, or .jpg files

\graphicspath{{figures/}{pictures/}{images/}{}} % where to search for the images

\usepackage{microtype}                 % use micro-typography (slightly more compact, better to read)
\PassOptionsToPackage{warn}{textcomp}  % to address font issues with \textrightarrow
\usepackage{textcomp}                  % use better special symbols
\usepackage{mathptmx}                  % use matching math font
\usepackage{times}                     % we use Times as the main font
\usepackage{cite}                      % needed to automatically sort the references
\usepackage{booktabs}                  % only used for the table example
\usepackage{subfig}
\usepackage{amsmath}
\usepackage{amssymb}
\usepackage{bm}

\usepackage{dsfont}

\usepackage{siunitx}
\newcommand{\revi}[1] {#1}

\newcommand{\revirm}[1] {}

\onlineid{1077}
\vgtccategory{Research}
\vgtcpapertype{Algorithm}

\title{Void-and-Cluster Sampling of Large Scattered Data and Trajectories}

\author{Tobias Rapp, Christoph Peters, and Carsten Dachsbacher}
\authorfooter{
 Tobias Rapp, Christoph Peters, and Carsten Dachsbacher are with Karlsruhe Institute of Technology. E-mail: \{tobias.rapp, christoph.peters, dachsbacher\}@kit.edu.
}

\abstract{
We propose a data reduction technique for scattered data based on statistical sampling. Our void-and-cluster sampling technique finds a representative subset that is optimally distributed in the spatial domain with respect to the blue noise property. In addition, it can adapt to a given density function, which we use to sample regions of high complexity in the multivariate value domain more densely. Moreover, our sampling technique implicitly defines an ordering on the samples that enables progressive data loading and a continuous level-of-detail representation.
We extend our technique to sample time-dependent trajectories, for example pathlines in a time interval, using an efficient and iterative approach.
Furthermore, we introduce a local and continuous error measure to quantify how well a set of samples represents the original dataset. We apply this error measure during sampling to guide the number of samples that are taken. Finally, we use this error measure and other quantities to evaluate the quality, performance, and scalability of our algorithm.
} % end of abstract


\CCScatlist{ % not used in journal version
 \CCScat{K.6.1}{Management of Computing and Information Systems}%
{Project and People Management}{Life Cycle};
 \CCScat{K.7.m}{The Computing Profession}{Miscellaneous}{Ethics}
}

\teaser{
  \centering
  \def\svgwidth{\columnwidth}
\begingroup%
  \makeatletter%
  \providecommand\color[2][]{%
    \errmessage{(Inkscape) Color is used for the text in Inkscape, but the package 'color.sty' is not loaded}%
    \renewcommand\color[2][]{}%
  }%
  \providecommand\transparent[1]{%
    \errmessage{(Inkscape) Transparency is used (non-zero) for the text in Inkscape, but the package 'transparent.sty' is not loaded}%
    \renewcommand\transparent[1]{}%
  }%
  \newcommand*\fsize{\dimexpr\f@size pt\relax}%
  \newcommand*\lineheight[1]{\fontsize{\fsize}{#1\fsize}\selectfont}%
  \ifx\svgwidth\undefined%
    \setlength{\unitlength}{562.69266464bp}%
    \ifx\svgscale\undefined%
      \relax%
    \else%
      \setlength{\unitlength}{\unitlength * \real{\svgscale}}%
    \fi%
  \else%
    \setlength{\unitlength}{\svgwidth}%
  \fi%
  \global\let\svgwidth\undefined%
  \global\let\svgscale\undefined%
  \makeatother%
  \begin{picture}(1,0.60043344)%
    \lineheight{1}%
    \setlength\tabcolsep{0pt}%
    \put(0,0){\includegraphics[width=\unitlength,page=1]{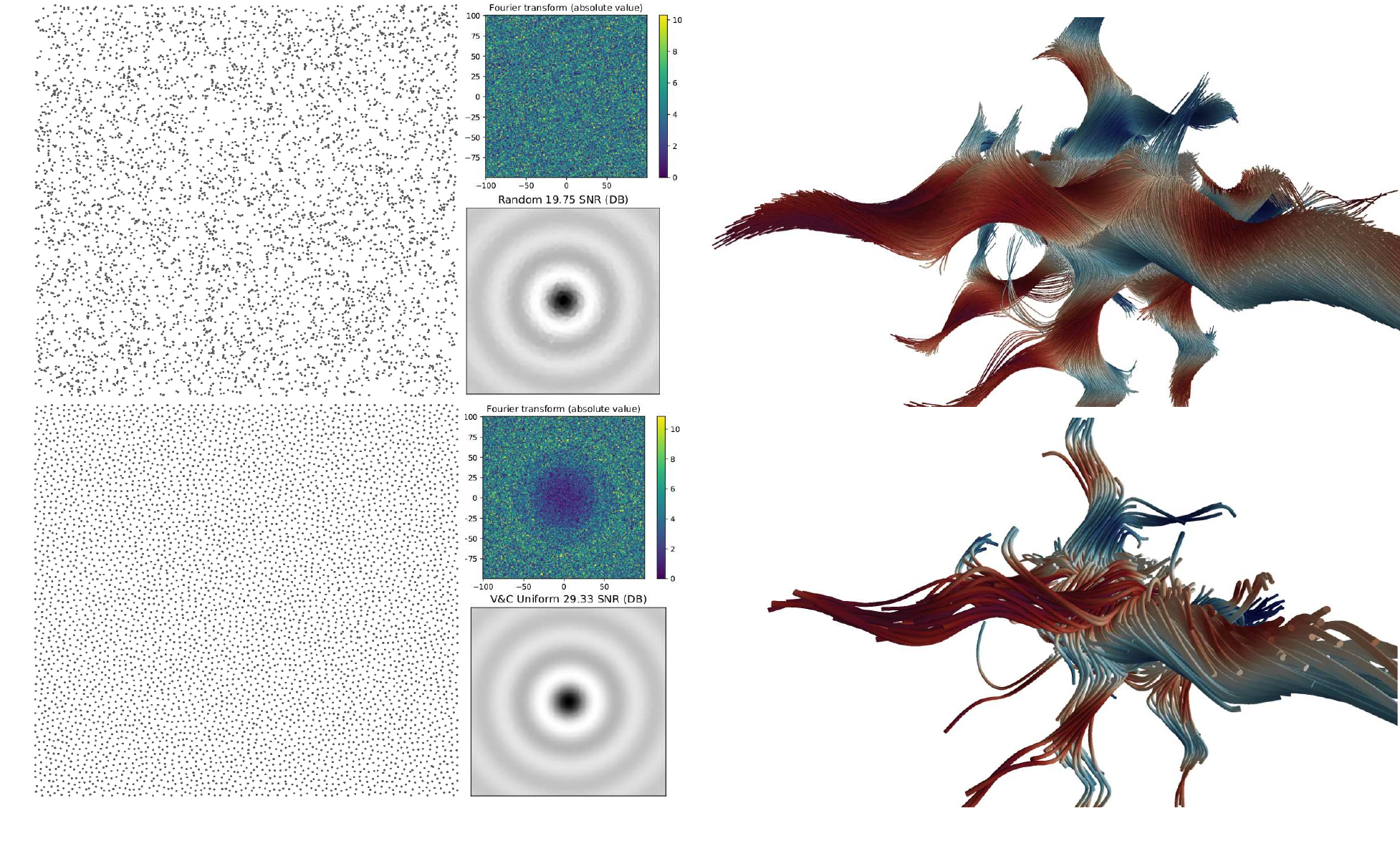}}%
    \put(-0.00170909,0.33434006){\color[rgb]{0,0,0}\makebox(0,0)[lt]{\begin{minipage}{0.1191572\unitlength}\raggedright (a)\end{minipage}}}%
    \put(-0.00079174,0.04891955){\color[rgb]{0,0,0}\makebox(0,0)[lt]{\begin{minipage}{0.1191572\unitlength}\raggedright (d)\end{minipage}}}%
    \put(0.96885544,0.33935937){\color[rgb]{0,0,0}\makebox(0,0)[lt]{\begin{minipage}{0.1191572\unitlength}\raggedright (g)\end{minipage}}}%
    \put(0.96664975,0.05632221){\color[rgb]{0,0,0}\makebox(0,0)[lt]{\begin{minipage}{0.1191572\unitlength}\raggedright (h)\end{minipage}}}%
    \put(0.47743507,0.4684972){\color[rgb]{0,0,0}\makebox(0,0)[lt]{\begin{minipage}{0.1191572\unitlength}\raggedright (b)\end{minipage}}}%
    \put(0.47739471,0.33003269){\color[rgb]{0,0,0}\makebox(0,0)[lt]{\begin{minipage}{0.1191572\unitlength}\raggedright (c)\end{minipage}}}%
    \put(0.47870732,0.18177009){\color[rgb]{0,0,0}\makebox(0,0)[lt]{\begin{minipage}{0.1191572\unitlength}\raggedright (e)\end{minipage}}}%
    \put(0.48141412,0.04804169){\color[rgb]{0,0,0}\makebox(0,0)[lt]{\begin{minipage}{0.1191572\unitlength}\raggedright (f)\end{minipage}}}%
  \end{picture}%
\endgroup%

  \caption{
  	We select \SI{1}{\percent} from \revi{$500,000$} data points by random (a) and using our void-and-cluster (d) sampling technique. Our approach chooses a set of samples that uniformly covers the spatial domain, whilst avoiding regularity artifacts, which leads to a blue noise spectrum for our technique (e) in contrast to random sampling (b). Our approach leads to a more accurate reconstruction of the dataset using the same amount of samples (c, f). Furthermore, we have sampled pathlines of the ABC flow with our technique (g, h), which implicitly defines a continuous level-of-detail, to render a greater (g) and smaller subset (h) of the trajectories.}
  \label{fig:teaser}
}

\newcommand \dd[1] { \,\textrm d{#1}}  % infintesimal
\newcommand{\SetFont}[1]{\mathbb{#1}}

\newcommand{\N}{\SetFont{N}}

\newcommand{\R}{\SetFont{R}}
\newcommand{\SampleSet}{S}
\newcommand{\SampleValueMap}{v}
\newcommand{\SampleValue}{V_S}

\newcommand{\Domain}{P}
\newcommand{\ValueDomain}{V}

\newcommand{\Entropy}{\mathbb{H}}
\newcommand{\Density}{\R^+}

\newcommand{\sden}{\lambda_S}
\newcommand{\pden}{\rho_P}
\newcommand{\pdenNon}{\tilde{\rho}_P}

\newcommand{\kernel}{k}
\newcommand{\kernelSize}{h_\Domain}
\newcommand{\kernelSizeSamples}{h}

\DeclareMathOperator*{\argmax}{arg\,max}
\DeclareMathOperator*{\argmin}{arg\,min}

\newcommand{\TightestClust}{s_{\max}}
\newcommand{\LargestVoid}{p_{\min}}

\newcommand{\Neighborhood}[1]{N_{#1}}

\newcommand{\iden}{\phi}
\newcommand{\HistogramBins}{N_{\text{bins}}}
\newcommand{\entden}{\phi_H}

\newcommand{\traj}{\tau}
\newcommand{\trajStart}{\traj_s}
\newcommand{\trajEnd}{\traj_e}
\newcommand{\TSamples}{T}
\newcommand{\TIntervalSize}{N}

\newcommand{\BatchMax}{\lambda_{\max}}
\newcommand{\Distance}[2]{\|p_{#1}-p_{#2}\|}

\newcommand{\ValueDistSamples}{\SampleValue}
\newcommand{\ValueDistPoints}{\ValueDomain}

\DeclareMathOperator{\Wasserstein}{W}

\newcommand{\CdfPoints}{{F_\ValueDomain}}
\newcommand{\CdfSamples}{{F_\SampleSet}}

\DeclareMathOperator{\sinc}{sinc}

\vgtcinsertpkg

\begin{document}

%% The ``\maketitle'' command must be the first command after the
%% ``\begin{document}'' command. It prepares and prints the title block.

%% the only exception to this rule is the \firstsection command
\firstsection{Introduction}
%% \section{Introduction} %for journal use above \firstsection{..} instead
\maketitle

In the field of scientific visualization, interactive exploration and analysis are considered essential to gain insight into large and complex datasets.
Although data sizes are growing rapidly, for example due to advancements in high-performance computing or increasingly accurate measurement devices,
storage bandwidth does not increase accordingly. Data reduction is thus a necessary means to reduce storage requirements for both simulation, measurement devices, and for subsequent data analysis.

We specifically consider the reduction of large, spatio-temporal scattered data\revirm{for visualization and analysis.}\revi{, i.e.\ unstructured points in space-time with an associated value domain.}
In particular, we investigate the use of statistical sampling to reduce large data sets to a representative subset. Sampling scales well to higher dimensional data and is well-suited for scattered data.
Although simple random sampling gives decent results, recent work improves upon this using stratified~\cite{Woodring2011, Su2014} and information-guided sampling~\cite{Wei2018, Biswas2018}. These results emphasize the significance of stratification in the spatial domain and adaptive sampling guided by the value domain. 

We propose a sampling strategy for scattered data \revirm{based on}\revi{generalizing} the void-and-cluster technique from Ulichney~\cite{Ulichney1993} that stratifies optimally in the spatial domain.
Specifically, we find samples that are well distributed with respect to the blue noise property, which implies large mutual distances between samples without causing regularity artifacts.
Additionally, we discuss how to adapt the sampling strategy to the value dimensions by better sampling regions of value distributions with high entropy. Moreover, the sampling technique implicitly defines an ordering on the samples that enables progressive data loading and continuous level-of-detail during visualization and analysis.
Our proposed algorithm is fast, scalable, and well-suited for GPU acceleration. Therefore, it is applicable in-situ, i.e.\ while a simulation is running, but also as a traditional post-processing step.

Furthermore, we extend our sampling technique to time-dependent scattered data. Instead of considering each time step independently, we sample trajectories, i.e.\ sequences of scattered points defined over time.
We find representative trajectories that evenly cover the spatio-temporal domain based on an efficient iterative extension of the void-and-cluster technique.
An example for such trajectory datasets are particle-based simulations that trace particles over time.
Additionally, representing fluid flows using Lagrangian trajectories, i.e.\ pathlines, instead of velocity fields has recently gained popularity~\cite{Agranovsky2014}. Pathlines are thereby advected during simulation time using the high-resolution vector field data, which could not be stored otherwise. In both of these examples, the data consists of a large amount of trajectories that we reduce using our sampling technique.

Lastly, we introduce an error measure to quantify how well a set of samples represents the data with respect to both the spatial and the value domain. In particular, we derive a continuous error measure that quantifies the difference in the value distributions for every point in the dataset.
This error measure integrates well into our sampling technique, where we use it to determine when a sufficient number of samples has been taken.
We evaluate the quality of our proposed sampling technique on different synthetic and real-word datasets using this error measure and other derived quantities, such as the quality of scattered data interpolation.
Finally, we investigate the performance and scalability of our proposed sampling technique and compare it to related approaches.

\noindent
To summarize, our contribution is a sampling technique that:
\begin{itemize}
\item Takes optimally distributed samples in the spatial domain with respect to the blue noise property,
\item Adapts to an arbitrary density, for example derived from a multivariate value domain,
\item Implicitly defines an ordering of the samples that we use for continuous level-of-detail and progressive data loading,
\item We extend to sample time-dependent data, for example pathlines in a fluid flow.
\end{itemize}

\section{Related Work}
\label{sec:Related}

We first discuss the visualization of large data with a focus on data reduction and scattered data, before we introduce the concept of blue noise and discuss corresponding sampling strategies.

\subsection{Visualizing Large Datasets}
To visualize large datasets, we focus on approaches that create a compact derived representation, instead of orthogonal approaches such as data compression. Li et al.~\cite{Li2018} survey data reduction techniques for simulation, visualization, and data analysis. 

Several distribution-based data representation approaches have been proposed, which represent large datasets using distributions that are sampled during subsequent visualization and analysis.
In particular, Thompson et al.~\cite{Thompson2011} represent value distributions by storing a histogram per block of voxels. Sicat et al.~\cite{Sicat2014} construct a multi-resolution volume from sparse probability density functions defined in the 4D domain comprised of the spatial and data range. Several promising approaches rely on Gaussian mixture models (GMMs), which represent arbitrary distributions as a weighted combination of Gaussians. Wang et al.~\cite{Wang2017} employ a spatial GMM in addition to a value distribution in each data block, whilst Dutta et al.~\cite{Dutta2017} partition the data into local, homogeneous regions and fit a GMM in each partition. For in-situ processing, Dutta et al.~\cite{Dutta2017a} perform incremental GMM estimation instead of expectation maximization to compute the mixture models.
Hazarika et al.~\cite{Hazarika2019} model distribution-based multivariate data using copula functions, which allow modeling the marginal distributions separately from the dependencies between dimensions.

Although distribution-based approaches achieve a significant reduction in data size, they are difficult to extend to higher-dimensional data due to the curse of dimensionality. Moreover, these approaches have been developed for uniformly structured data and the extension to scattered data is non-trivial. While scattered data can be visualized by first reconstructing a structured representation~\cite{Fraedrich2010, Reichl2013}, this approach has its own drawbacks and is not an option for all analysis techniques and needs. For example, particle-based visualizations benefit from specific visualization and analysis techniques~\cite{Grottel2014}.

\subsection{Sampling}
Statistical sampling of data~\cite{Dix2002} is gaining popularity in the field of scientific visualization. Reinhardt et al.~\cite{Reinhardt2017} use stochastic sampling to improve performance and reduce visual clutter for the visual debugging of smoothed particle hydrodynamics (SPH) simulations.
Sauer et al.~\cite{Sauer2017} propose a data representation that combines particle and volume data and supports sampling of particles by using the corresponding volume to find evenly distributed samples.
Woodring et al.~\cite{Woodring2011} describe a simulation-time stratified sampling strategy for a large-scale particle simulation. For stratification, the authors construct a kd-tree from the data that is also used as a level-of-detail representation.
Su et al.~\cite{Su2014} discuss server-side sampling using bitmap indices and stratify both in the spatial and value domain.
Wei et al.~\cite{Wei2018} extend their approach with an information-guided sampling strategy and recovery technique. During sampling, they measure the information per stratum by computing the entropy of the value distribution and draw samples accordingly.
Biswas et al.~\cite{Biswas2018} similarly employ an information-guided strategy to sample adaptively during in-situ simulation, but use a global histogram for the entropy computation.

A desirable property of point distributions is the blue noise characteristic~\cite{Ulichney1988}, which leads to large mutual distances between points without noticeable regularity artifacts.
Balzer et al.~\cite{Balzer2009} compute Capacity-constrained Voronoi diagrams (CCVD) to optimize the blue noise property of point distributions, which allows adapting the point distributions to a given density function.
To find representative particles, Frey et al.~\cite{Frey2011} propose loose capacity-constrained Voronoi diagrams (LCCVD) that relax the capacity-constraints of the CCVD method and are computed on the GPU\@. All methods based on capacity-constrained Voronoi diagrams can be used to sample scattered data, but are computationally demanding.
\revi{Bridson~\cite{Bridson2007} presents a Poisson disk sampling technique to generate blue noise samples in arbitrary dimensions by enforcing a minimal and maximal distance between nearest neighbors. The technique is designed to produce entirely new sample sets, not to reduce an existing sample set.}

\begin{figure*}[t]
	\centering
	\subfloat[Initial random sampling]{%
		\includegraphics[height=2.75cm]{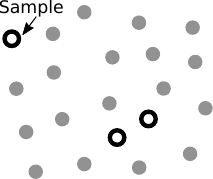}
	}
	\hfill
	\subfloat[Find $\LargestVoid$ and $\TightestClust$]{%
		\includegraphics[height=2.75cm]{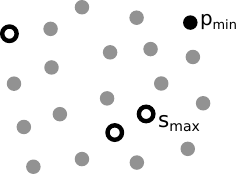}
	}
	\hfill
	\subfloat[Exchange $\LargestVoid$ and $\TightestClust$]{%
		\includegraphics[height=2.75cm]{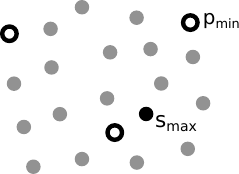}
	}
	\hfill
	\subfloat[Find and add the next $\LargestVoid$]{%
		\includegraphics[height=2.75cm]{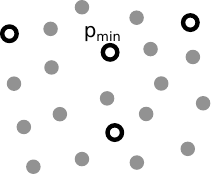}
	}
	\caption{Overview of the void-and-cluster sampling technique of Ulichney\revirm{ et al.}~\cite{Ulichney1993} extended to scattered data. After initial random sampling (a), the samples are optimized by finding (b) and exchanging (c) the largest void $\LargestVoid$ with the tightest cluster $\TightestClust$ until $\LargestVoid = \TightestClust$. We then iteratively find and add (d) the largest void $\LargestVoid$ until we have enough samples.}
	\label{fig:Overview}
\end{figure*}

Ulichney~\cite{Ulichney1993} introduces the void-and-cluster sampling technique in the context of halftoning and dithering.
The technique ranks all pixels in a rastered image, thus producing a dithering mask. If we think of all pixels that are already ranked as white and mark the others black, applying a Gaussian filter yields an image that indicates the local density of ranked pixels. The tightest cluster is bright\revi{est}, the largest void is dark\revi{est}. The void-and-cluster technique goes through three phases to fill large voids and reduce tight clusters greedily. \revi{The order of greedy additions implies an order on the sample set such that each prefix has good blue noise characteristics.}\revirm{In the end, all pixels are ranked and the resulting dithering patterns have blue noise characteristics.}
\revirm{In the next sections, we discuss the extension of this technique to scattered data, augment it with non-uniform sampling densities, and propose a parallel implementation.}

\section{Sampling Scattered Data and Trajectories}
\label{sec:Scattered}

\revirm{In this section, we introduce the void-and-cluster sampling technique for scattered data, extend it to non-uniform sampling probabilities, and show how to sample time-dependent trajectories.}
\revi{
Ulichney's algorithm~\cite{Ulichney1993} is restricted to regular grids and produces samples with a uniform density. 
In this section, we generalize the approach to scattered data with a non-uniform distribution. 
To preserve the spatial density, we compute a density estimate on the whole dataset. 
Our generalized void-and-cluster sampling works on scattered data and enforces the given density (\autoref{sec:Scattered:Algorithm}). 
Like the original algorithm, it orders all sample points to enable level of detail and progressive data loading (\autoref{sec:Scattered:Ordering}). 
The algorithm is efficient because each iteration only requires local updates with compact kernels (\autoref{sec:Scattered:Kernels}).
Our parallel implementation in \autoref{sec:Parallel} further exploits this locality. 
Supporting arbitrary sampling densities lets us emphasize regions of high entropy in the value domain (\autoref{sec:Scattered:Densities}).
Finally, we extend our technique to the sampling of time-dependent trajectories (\autoref{sec:Scattered:Trajectories}).
}

\subsection{Void-and-Cluster Sampling}
\label{sec:Scattered:Algorithm}

Assume we have a dataset with points $\Domain\subset\R^d$ in a $d$-dimensional spatial domain. Each sample is mapped to a value in the possibly multivariate value domain $\ValueDomain$ through $\SampleValueMap: \Domain\rightarrow\ValueDomain$. Among these points, we want to pick a representative subset $\SampleSet \subset \Domain$.
Therefore, we optimize the placement of samples in the spatial domain by estimating the density of selected samples $\sden: \Domain \rightarrow \Density$ for each point $p \in \Domain$. A high sample density indicates a large number of nearby samples, whilst a low density indicates few.
We want to place samples such that dense regions (clusters) and empty regions (voids) are avoided. Or, in other words, reduce the maximum of the sample density $\sden$ and increase its minimum.
 
This does not work\revirm{, however,} for spatially non-uniformly distributed data points since we have to account for the original distribution in the spatial domain.
Even for uniformly distributed points the border region of the spatial domain is less densely populated.
We account for the spatial distribution of the points by first computing a point density $\pden$ for each $p \in \Domain$:
\begin{equation}
\pden(p) := \sum_{p_i \in \Domain} \kernel(\|p - p_i\|),
\end{equation}
using a kernel function $\kernel$. Given a subset of samples $\SampleSet\subset\Domain$, the sample density at $p\in\Domain$ is then defined as:
\begin{equation}
	\sden(p) := \frac{\sum_{s \in \SampleSet} \kernel(\|p - s\|)}{\pden(p)}.
\end{equation}
We will now describe a strategy to find the optimal set of samples in the spatial domain with respect to the sample density $\sden$, by extending the void-and-cluster algorithm~\cite{Ulichney1993}. This is an iterative and greedy algorithm that at each step finds a locally optimal distribution of samples. An overview of our modified void-and-cluster sampling technique is depicted in \autoref{fig:Overview}. 

Initially, we take a fixed number of random samples.
Although the point density $\pden$ stays constant, we have to update $\sden$ when we change the set of samples. The sample density is computed incrementally when a sample $s$ is added (or removed), by adding to (or subtracting from) the density $\sden(p)$ for all points.

We then optimize these initial samples by removing the tightest cluster
\begin{equation}
\TightestClust = \argmax_{s \in \SampleSet}\sden(s)\in\SampleSet,
\end{equation}
i.e.\ the sample with largest $\sden$.
Then, we add the largest void
\begin{equation}
\LargestVoid = \argmin_{p \in \Domain \setminus \SampleSet}\sden(p)\in\Domain \setminus \SampleSet,
\end{equation}
i.e.\ the point with the lowest $\sden$ that is not a sample yet. Since we add and remove a sample, we have to update the sample densities accordingly. The optimization stops once the tightest cluster that we remove then becomes the largest void, that is $\TightestClust = \LargestVoid$.

After construction of the optimal initial sampling, we iteratively find and add the largest void to the set of samples until we have reached the desired amount of samples. 
We provide detailed pseudocode of the entire algorithm in the supplementary document.

\subsection{Ordering of Samples}
\label{sec:Scattered:Ordering}

A positive side-effect of the greedy approach is that the void-and-cluster strategy implicitly defines an ordering of the samples. With respect to this ordering, any prefix of the sample set $\SampleSet$ still has good blue noise characteristics. Ulichney~\cite{Ulichney1993} denotes it as the rank \revi{$r: \Domain \rightarrow \N$, where $r(p) = \infty$ for all $p \in \Domain\setminus\SampleSet$}.
To compute this ordering, we assign and increment the rank when adding a sample during the initial random sampling or the void filling steps. For a sample $s_i$ that is added as the $i$-th sample, we set $r(s_i) = i$. During the void-and-cluster optimization, when we exchange the tightest cluster \revi{$\TightestClust$} with the largest void \revi{$\LargestVoid$}, we have to swap the rank accordingly\revi{, i.e.\ we set $r(\LargestVoid) = r(\TightestClust)$ and $r(\TightestClust) = \infty$}.
%
%}\revi{we set the rank of the largest void equal to the tightest cluster: $r(\LargestVoid) = r(\TightestClust)$}.

We re-order (or index) the samples according to this mapping.
We can use this ordering for continuous level-of-detail and for progressive data loading during the subsequent visualization and analysis.

\subsection{Compact Kernels}
\label{sec:Scattered:Kernels}

If the kernel $\kernel$ is compact, i.e.\ has a finite extent, only a local neighborhood has to be considered when updating the densities of samples and points. For compact kernels, the optimization is thus defined locally. In our experiments, we found that the choice of kernel does influence the distribution of samples and the quality of the blue noise. Nonetheless, we always achieved good results as long as the kernel size $\kernelSize$ was in a reasonable order of magnitude with respect to the spatial domain.
If we take a fraction of all samples $|\SampleSet|<|\Domain|$, we have to increase the kernel size $\kernelSizeSamples$ used for sampling accordingly:
\begin{equation}
	\kernelSizeSamples := \kernelSize\sqrt[d]{\frac{|\Domain|}{|\SampleSet|}},
\end{equation}
using the spatial dimension $d$.
Although we did experiment with a Gaussian kernel, we use a cubic spline~\cite{Monaghan1992} in the remainder of this work since it yields similar results, but is compact. Lastly, we denote points in the support of kernel $\kernel$ at point $p \in \Domain$ as its neighborhood $\Neighborhood{p}\subset\Domain$.

\subsection{\revi{Adaptive} Sampling}
\label{sec:Scattered:Densities}

\revirm{For quantitative analysis and visualization using a subset of a dataset, the sampling must be representative, i.e.\ accurately represent the original dataset.}
\revi{So far we have taken all samples with equal probability and proportional to the spatial density. Now, we discuss the use of} non-uniform probabilities to better capture complicated behavior in the value dimensions.

In general, we would like to take samples $\SampleSet \subset \Domain$ according to the probability mass function $\iden: \Domain \rightarrow [0,1]$. To sample a representative subset, we must re-weight all samples $s \in \SampleSet$ proportionally to the reciprocal $\iden^{-1}(s)$.
With our void-and-cluster approach, we implement this adaptation by using a modified density:
\begin{equation}
	\pdenNon(p) := \pden(p) \iden(p).
\end{equation}
\revi{Thus, any normalized importance measure, for example a computed feature or derived variable, can be used to guide the placement of samples.}

\paragraph{Entropy Sampling}

\begin{figure}
	\includegraphics[width=0.9\linewidth]{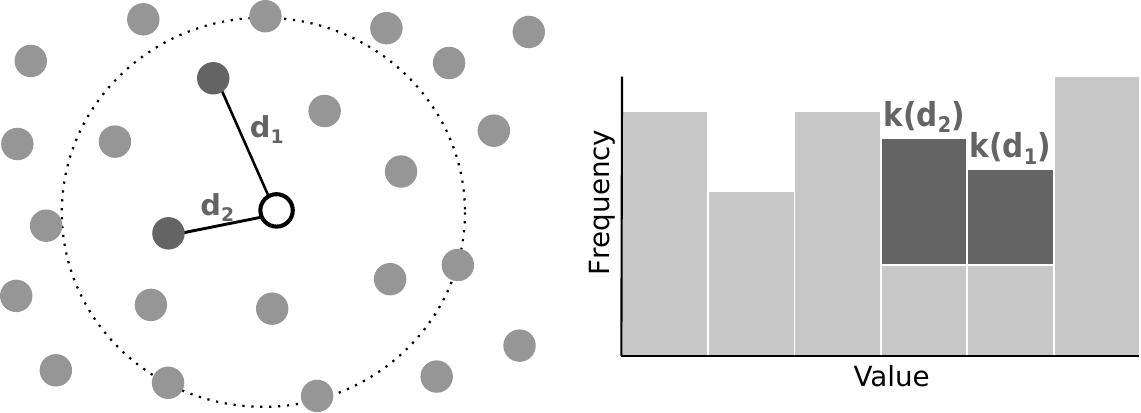}
	\caption{We compute the entropy of a point in its local neighborhood from a histogram of the value distribution, weighted by the radially symmetric kernel $\kernel$. }
	\label{fig:Entropy}
\end{figure}

Similar to recently proposed sampling techniques~\cite{Wei2018, Biswas2018}, we place more samples in regions with a high entropy, i.e.\ value distributions of high complexity.
 
For each point $p \in \Domain$ we compute the entropy using its local neighborhood $\Neighborhood{p}$, see \autoref{fig:Entropy}. Specifically, we create a histogram of the value distribution of all points in the neighborhood. \revi{We use the global value range for the computation of the histogram to ensure that the entropy is consistent everywhere.} To obtain a continuous entropy in the spatial domain, we weight the contribution of each neighbor with respect to its distance to $p$ using the kernel $\kernel$. From the weighted and normalized histogram $h_p$ of size $\HistogramBins$, we compute the entropy:
\begin{equation}
\Entropy(p) := - \sum_{i = 0}^{\HistogramBins - 1} h_p(i) \log_2 h_p(i).
\end{equation}
Similar to Wei et al.~\cite{Wei2018}, we derive a sampling probability that is independent of the size of the histogram as
\begin{equation}
	\entden(p) := \frac{2^{\Entropy(p)}}{N_{\text{bins}}}
\end{equation}
and then derive a correctly normalized probability mass function as
\begin{equation}
	\iden(p) = \frac{\entden(p)}{\sum_{p_i \in \Domain}\entden(p_i)}.
\end{equation}
For multivariate data, we have to construct a single \revi{probability} from multiple value dimensions. Hence, we compute the entropy individually in each dimension and use the maximal entropy at each point.
\revi{Intuitively, we consider a data point relevant if at least one dimension shows high entropy. Dependent on the application}, we could \revi{also select} a subset of the value dimensions \revi{to guide the entropy sampling.}

\subsection{Trajectory Sampling}
\label{sec:Scattered:Trajectories}

In addition to sampling a single time step, we extend our sampling technique to time-dependent data.
Specifically, we consider trajectories of scattered data points in discrete time steps $t_0, ..., t_{\TIntervalSize-1} \in \R$. A trajectory is then defined as a sequence of points over time $\traj := (p_{t_j}, \dots, p_{t_k})$ with $0 \leq j \leq k \leq \TIntervalSize - 1$ and points $p_{t_i} \in P_{t_i}$ at time $t_i$. Note that each trajectory can have different starting and ending points in time, i.e.\ it does not have to be present in every time step. We now discuss how to sample a subset $\TSamples$ from the set of all trajectories.

To avoid an optimization of trajectories over all time steps, we sample iteratively. In the first time step, we employ our void-and-cluster sampling strategy to sample a subset $S_{t_0} \subset P_{t_0}$ that defines an initial set of trajectories $\TSamples$.
In the next time step, a number of trajectories could end, i.e.\ no longer exist in the following steps. We first compute the point density $\pden$ and the sample density $\sden$ from the trajectories $\TSamples$ that still exist in the current time step.
For $n$ ending trajectories, we then add the trajectories from the $n$ largest voids to $\TSamples$ and thus start new trajectories from this time step. To start a trajectory $\traj$ in time step $t_i$ means that we create a new trajectory $\trajStart := (p_{t_i},\ldots,p_{t_k})$.

\begin{figure}
	\includegraphics[width=\linewidth]{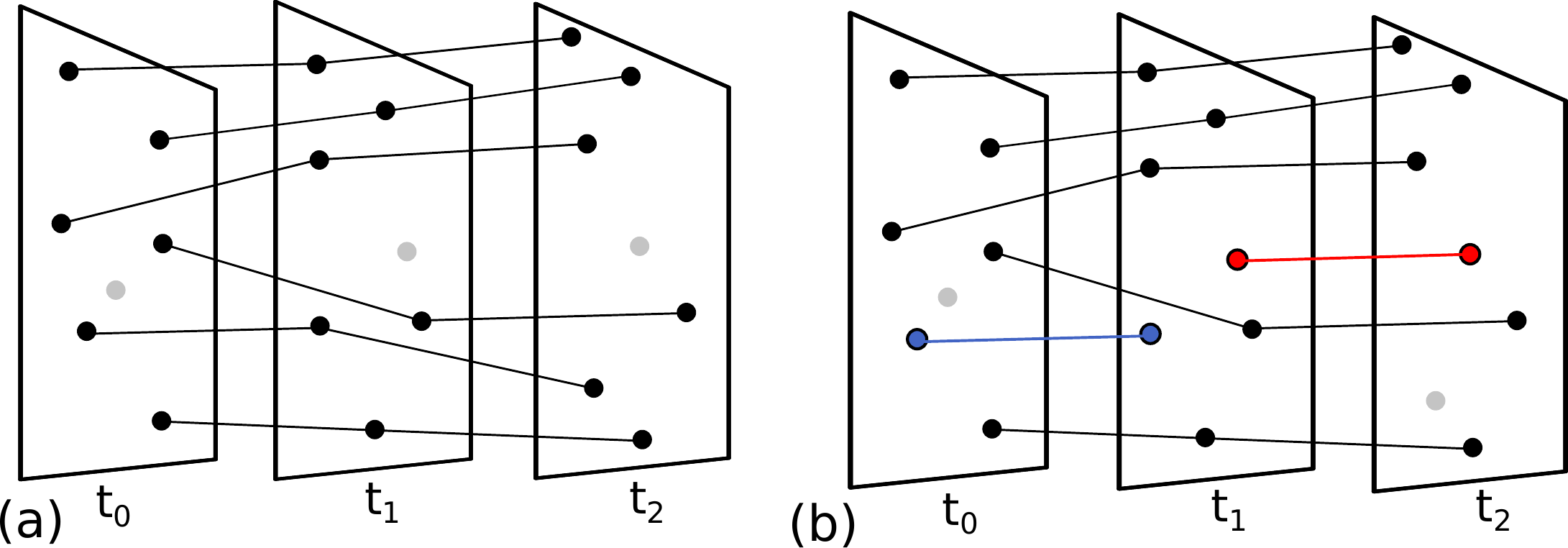}
	\caption{In (a), we sample trajectories that bundle and separate over time. We optimize the \revirm{sample} distribution \revi{of sampled trajectories} in \revirm{each step} (b), by stopping (blue) and starting (red) trajectories in $t_1$.}
	\label{fig:TrajectorySampling}
\end{figure}

Hlawatsch et al.~\cite{Hlawatsch2011} observed that longer trajectories have greater accuracy than a series of shorter trajectories\revi{. However,} longer trajectories may bundle together or move away and create regions with little coverage, see \autoref{fig:TrajectorySampling}. Thus, we forcefully stop up to a user-defined amount of trajectories $\epsilon_T$ in each time step $t_i$. To stop a trajectory $\traj$ in time step $t_i$, we take the prefix $\trajEnd := (p_{t_j},\ldots,p_{t_i})$ instead of $\traj$.
The parameter $\epsilon_T$ depends on the dataset \revi{and the specific application. In general, it} should be inversely proportional to the amount of trajectories ending. In datasets where all trajectories exist in all time steps, $\epsilon_T$ should be high.
To select which trajectories to stop and which to start in time step $t_i$, we perform the void-and-cluster optimization. That is, we exchange the tightest cluster with the largest void up to $\epsilon_T$ times or until the sample distribution is optimal, i.e.\ the tightest cluster is equal to the largest void, and start or stop the corresponding trajectories. \revi{Note that longer trajectories may again be obtained by interpolation from shorter ones~\cite{Agranovsky2014}.} 

\section{Parallel Implementation}
\label{sec:Parallel}

In this section, we discuss the parallel implementation of the sampling technique, specifically the computation of the point and sample densities, and the parallelization of the void filling step.

\subsection{Computing the Densities}
One of the most computationally demanding parts of the algorithm is creating the density $\pden$ and updating $\sden$. Each point $p$ has to scatter its density, weighted by the distance and kernel function, to all neighboring points $\Neighborhood{p}$. This is an embarrassingly parallel task and is especially well-suited for GPU acceleration. If the kernel function is compact, data structures, such as a kd-tree or a regular grid, should be employed to efficiently retrieve the neighborhood of a point or sample.

In our implementation, we use a uniform grid of cell size $\kernelSizeSamples$. To find the neighborhood $\Neighborhood{p}$ of $p$, we thus have to query $3^d$ cells around $p$. To speed-up the neighborhood search, we layout all cells in memory using a space-filling Z-curve to optimize memory access to neighboring cells.

\subsection{Parallel Void Filling}
The void filling step seems to enforce a sequential bottleneck: In each step, we find the sample $p\in\Domain\setminus\SampleSet$ with the smallest sample density $\sden(p)$. Then we add $p$ to the sample set $\SampleSet$ and increase sample densities in the neighborhood before we search the smallest sample density again. To overcome this sequential dependency, we store each added sample $p$ alongside the sample density $\sden(p)$ that it has when it is added. Since we only ever add to the densities and pick the minimum in each step, these densities grow monotonically. If we can guarantee that they are computed correctly for each added sample, sorting by the densities guarantees that we rank all selected samples correctly.

To provide this guarantee, we must never add a sample too early. All samples in a neighborhood $p_n\in\Neighborhood{p}$ with a rank $r(p_n)<r(p)$ must have contributed to the sample density $\sden(p)$ before we add $p\in\Domain\setminus\SampleSet$ and store $\sden(p)$. We can be certain that this is the case if $p$ has the smallest sample density in its neighborhood, i.e.
\begin{equation}
	\sden(p) \leq \min_{p_n \in \Neighborhood{p}}\sden(p_n).
	\label{eq:SmallestDensityInNeighborhood}
\end{equation}
By selecting the sample $p\in\Domain\setminus\SampleSet$ that minimizes $\sden(p)$ globally, we will never select another sample in $\Neighborhood{p}$ before $p$. Hence, we know that the rank $r(p)$ is also minimal within the neighborhood $\Neighborhood{p}$.

This principle enables our parallel implementation. We first sort all $p \in  \Domain \setminus \SampleSet$ in ascending order by their density $\sden(p)$ and take the first $n$ points $p_0, p_1, \dots, p_{n-1}$ in each iteration. Then we compute an adjacency matrix in parallel using the kernel size $\kernelSizeSamples$:
\begin{equation*}
A := 
\begin{pmatrix}
0      & 0 &  \dots & 0 \\
\Distance{1}{0} - \kernelSizeSamples & 0 & \dots & 0  \\
\vdots & \ddots & \ddots & \vdots \\
\Distance{n-1}{0} - \kernelSizeSamples & \dots & \Distance{n-1}{n-2} - \kernelSizeSamples & 0 \\
\end{pmatrix}.
\end{equation*}
A negative entry in the $i$-th row and $j$-th column with $i>j$ indicates that $p_i$ is in the neighborhood of $p_j$. Since $\sden(p_i)\ge\sden(p_j)$ due to the sorting, this means that $p_i$ might not satisfy Equation~(\ref{eq:SmallestDensityInNeighborhood}) and it is flagged accordingly. Once the process is complete, all points that have not been flagged satisfy Equation~(\ref{eq:SmallestDensityInNeighborhood}). They are added to the sample set and their densities are stored. Note that the matrix $A$ is never stored. We only need the flags.

Although we can parallelize this computation, the workload is unevenly distributed. The $i$-th row of $A$ has $i$ non-zero entries. Therefore, we index the non-zero entries of $A$ with a single linear index $k \in \{0,\ldots,\frac{n(n-1)}{2}\}$. In the supplementary material we show that row and column indices can be computed from this flat index through
\begin{equation}
i = \left\lfloor\frac{1}{2} + \sqrt{\frac{1}{4} - 2k}\right\rfloor, \qquad
j = k - \frac{(i - 1)i}{2}.
\end{equation}
For large $k$, the square root has to be evaluated in double-precision to avoid rounding errors.

Thanks to the sorting by density, the procedure described above guarantees a correct relative rank of selected samples. However, samples may be missing if we just terminate after a particular iteration. For a complete result, we perform additional iterations. If the largest density of a sample added in the last proper iteration was $\BatchMax$, we continue iterating until the smallest sample density in $\Domain\setminus\SampleSet$ is greater than $\BatchMax$. At this point, we can be certain that we have not missed a sample that should have been added up until the last proper iteration. This way, we guarantee that we take the same samples as the sequential algorithm.

\section{Local Error Measure}

In this section, we discuss an error measure to quantify how well a set of samples represents a dataset. We propose a measure that takes not only the spatial domain into account, but also how well the value domain is represented in each region of the dataset. To this end, we first discuss how such a local error can be defined, before we discuss how to compare value distributions. Lastly, we discuss an error guided sampling strategy that relies on an efficient iterative error estimation to sample just below a given error threshold, instead of drawing a fixed amount of samples.

\subsection{Locality and Continuity}

We derive a local error measure that compares the value distribution $\ValueDistSamples \subset \ValueDomain$ of the sampled dataset with the value distribution $\ValueDomain$ of the original dataset. Specifically, we propose to compare value distributions in the local neighborhood $\Neighborhood{p}$ for each corresponding $p \in \Domain$. This method implicitly accounts for non-uniformly distributed data points. Additionally, we weight the contribution of each $p_i \in \Neighborhood{p}$ to the value distribution by its distance $\kernel(\|p - p_i\|)$ so that the error varies smoothly over the spatial domain.

\subsection{Wasserstein Distance}

To measure the difference between the original value distribution given by values $\ValueDistPoints = \{X_0, ..., X_{n-1}\}$ and a sampled subset $\ValueDistSamples \subset \ValueDistPoints$, we use the corresponding cumulative distribution functions (CDFs) $\CdfPoints$ and $\CdfSamples$. The CDF \revi{at a point $p \in \Domain$} is estimated as
\revi{
\begin{equation}
\CdfPoints(p, t) = \frac{1}{\sum_{i=0}^{n-1} \kernel(\|p - p_i\|)} \sum_{i=0}^{n-1}
\begin{cases}
\kernel(\|p - p_i\|) & \text{if } X_i \leq t, \\
0 & \text{otherwise}.
\end{cases}
\end{equation}
}
In practice, this implies that we need to sort the $X_i$ before accumulating them. Since the samples are a subset $\ValueDistSamples \subset \ValueDistPoints$, it is sufficient to sort the values $\ValueDistPoints$ to estimate both CDFs. Alternatively, we estimate the CDFs based on a histogram of $\ValueDistPoints$ and $\ValueDistSamples$, which introduces a discretization, but is more efficient to evaluate.

To measure the distance, we found the Wasserstein distance, or earth movers distance, to be a good choice. In the one-dimensional case, it is defined as the L1-norm between the two CDFs:
\revi{
\begin{equation}
	\Wasserstein(p, \CdfPoints, \CdfSamples) := \int_{-\infty}^{\infty} \left| \CdfPoints(p, x) - \CdfSamples(p, x)\right| \dd{x}.
\end{equation}
}
In contrast, we found the Kolmogorov-Smirnov distance, defined as the infinity norm between the CDFs, to be unsuited since it is not robust to small shifts in the value dimension.

Note that this definition of the Wasserstein distance is only valid for one-dimensional value distributions. Thus, we compute a separate error for each value dimension. We can further deduce the error across dimensions, e.g.\ by taking the mean or maximum. In the following we will use the maximum\revi{;} however, this is an application and data specific decision.

\subsection{Error Guided Sampling}

During void-and-cluster sampling, we efficiently keep track of the error distribution, for example to stop sampling if the average error falls below a given threshold.
In detail, we compute the error for all samples after the initial void-and-cluster optimization. When adding a sample $\LargestVoid$, we compute the error of $\LargestVoid$ and additionally update the error for all neighbors $\Neighborhood{\LargestVoid}$ since these have changed as well.

To describe the distribution of errors during sampling, we found the average error to be a robust statistic that is efficient to compute. In contrast, the maximal error does not decrease smoothly with respect to the number of samples and is not robust against outliers, e.g.\ stemming from small, but complex value regions.

\section{Results and Discussion}
\label{sec:Results}

In this section, we evaluate our sampling technique using \revi{four} real-world datasets and the synthetic $\sinc$ signal.

\subsection{Synthetic Data: Sinc}

\begin{figure*}
	\includegraphics[width=\linewidth]{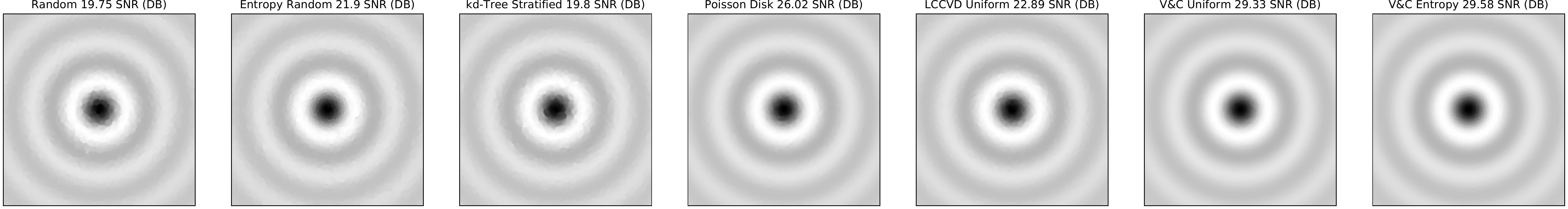}
	\caption{Reconstruction of the $\sinc$ dataset using scattered data interpolation after taking $5,000$ samples with different strategies.}
	\label{fig:sinc:rec}
	\vspace*{-2mm}
\end{figure*}

\begin{figure}
	\includegraphics[width=0.49\linewidth]{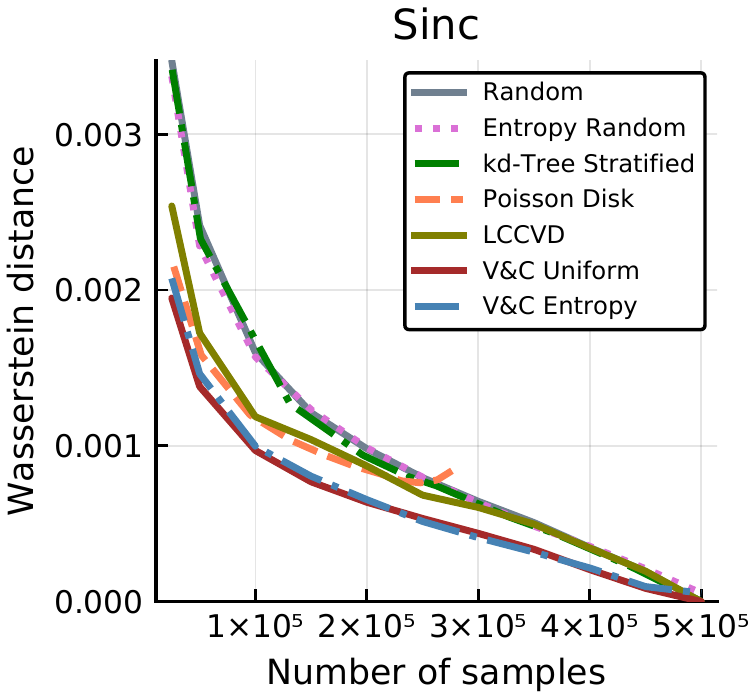}%
	\includegraphics[width=0.49\linewidth]{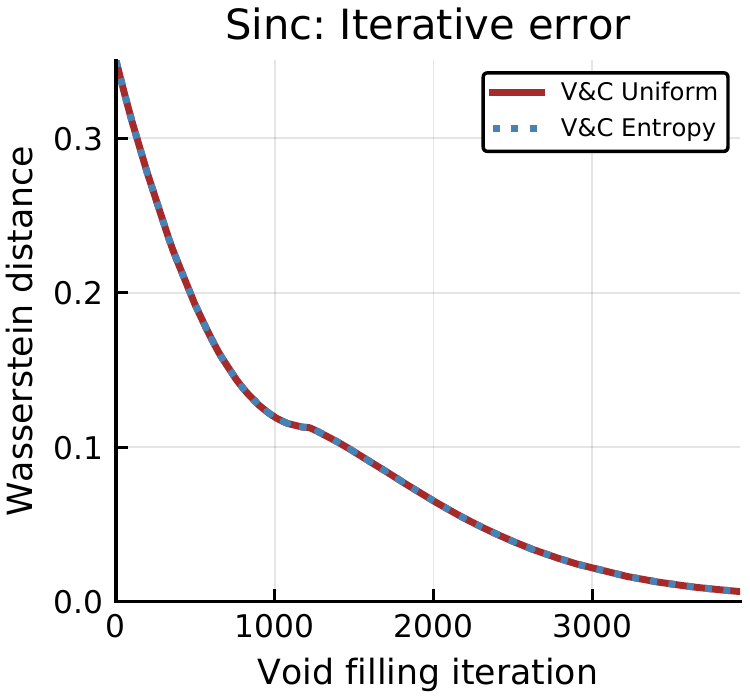}
	\vspace*{-2mm}
	\caption{Left: Comparison of different sampling strategies using our proposed error measure. Right: Error measured during sampling.}
	\label{fig:sinc:error}
\end{figure}

\begin{figure}
	\includegraphics[width=0.99\linewidth]{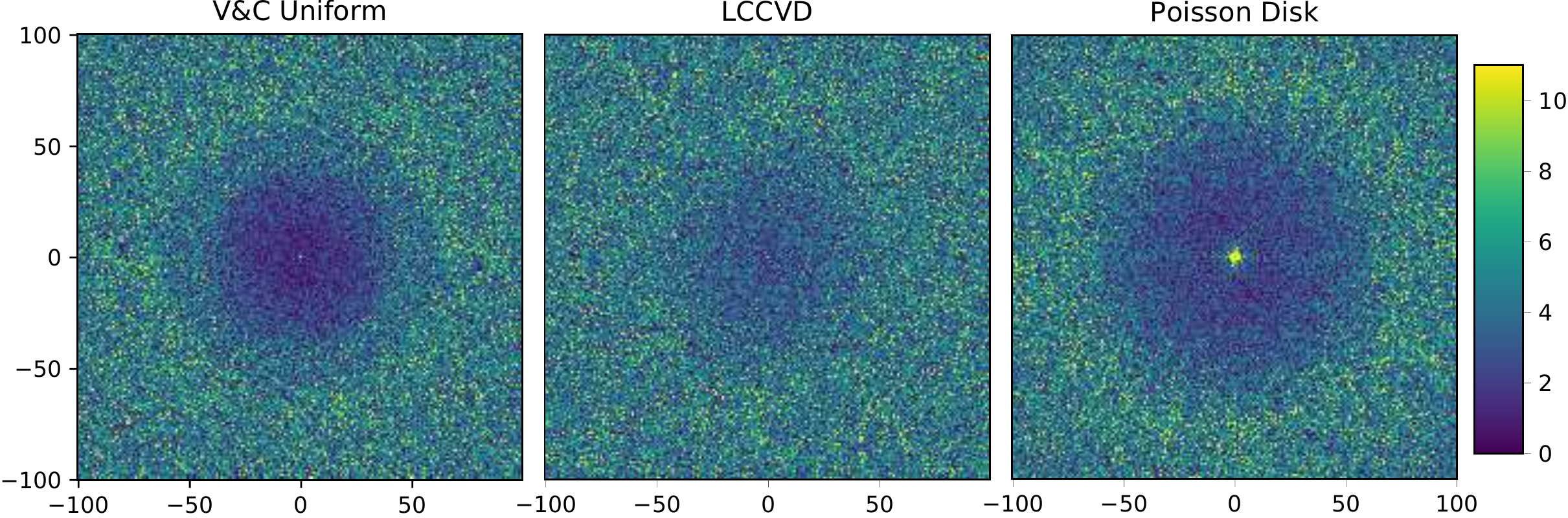}
	\caption{Fourier transform of the $\sinc$ dataset after taking $5,000$ samples using our uniform void-and-cluster method\revirm{ (left) and using}\revi{,} loose capacity constrained Voronoi diagrams (LCCVD)\revirm{ (right)}\revi{, and Poisson disk sampling}.}
	\label{fig:sinc:spectra}
	\vspace*{-1mm}
\end{figure}

We have created the $\sinc$ dataset by randomly placing $500,000$ points in the domain $[-5, 5]^2$ and by evaluating for each point $p \in [-5, 5]^2$ the function 
\begin{equation}
	\sinc(\|p\|) = \frac{\sin(\pi \|p\|)}{\pi \|p\|}.
\end{equation}
\autoref{fig:sinc:error} (left) compares different sampling strategies and shows the mean Wasserstein distance. We employ simple random sampling and random sampling with non-uniform probabilities based on the entropy. Moreover, we compare to stratified sampling utilizing a kd-tree based on a median split, similar to Woodring et al.~\cite{Woodring2011}.
Lastly, we employ \revi{Poisson disk sampling~\cite{Bridson2007} and} loose capacity constrained Voronoi diagrams (LCCVD,~\cite{Frey2011}).
For \revi{an} increasing \revi{number of samples}, the error from most strategies converges to zero, which implies that \revi{these} strategies sample a representative subset.
\revi{However, Bridson's Poisson disk sampling~\cite{Bridson2007} lacks explicit control over the sample count, which is instead steered by the enforced minimal and maximal distance. The technique is unable to surpass a certain sample count for this dataset.}
\revi{O}ur proposed void-and-cluster sampling strategies perform \revirm{consistently better} \revi{best for all sample counts}. The entropy-based strategies perform similar to their uniform counter parts. 

In \autoref{fig:sinc:error} (right), the mean error has been computed iteratively during sampling until the error was less than $\epsilon = 0.0065$, which led to a sampling percentage of \SI{34.1}{\percent}. The error first falls rapidly then converges asymptotically to zero.
Initially, we sample $5,000$ and iteratively add the remaining samples. In each void filling step, we take only $32$ samples in parallel to still keep the error up to date. In comparison, we take up to $12,288$ voids in parallel if we do not perform error guided sampling.

We use scattered data interpolation to interpolate the sampled data values to a grid of size $1024^2$, see \autoref{fig:sinc:rec}. Our void-and-cluster strategies show a major improvement compared to the other sampling strategies. The signal-to-noise (SNR) ratios shown in the logarithmic decibel scale support this assessment. Note that the LCCVD \revi{and Poisson disk} sampling strateg\revi{ies} also achieve\revirm{s} good results, but \revi{are} still worse than our proposed methods. For reconstruction, the entropy-based sampling strategies perform slightly better compared to the uniform approaches. For all sampling strategies, the quality of the reconstruction agrees with the error measure.

Lastly, \autoref{fig:sinc:spectra} shows the spectrum of the $\sinc$ dataset reduced to a subset of \SI{1}{\percent} with our uniform void-and-cluster strategy\revirm{ (left) and with LCCVD (right)}\revi{, LCCVD, and with Poisson disk sampling}. The Fourier transform shows the blue noise property for \revi{these} methods. Low frequencies are substantially weaker and the spectrum is isotropic. However, LCCVD has more energy in low frequencies than our void-and-cluster strategy. \revi{Poisson disk sampling has less energy in low frequencies, but contains a noticeable spike near zero.} Note that the random and stratified sampling strategies do not have this property\revi{, which suggests that the blue noise property is desirable for scattered data interpolation. Indeed, the error of kernel estimation has been shown to depend on the disorder of particles~\cite{Monaghan1982}}.

\subsection{Von K\'{a}rm\'{a}n Vortex Street}

\begin{figure}
	\includegraphics[width=0.97\linewidth]{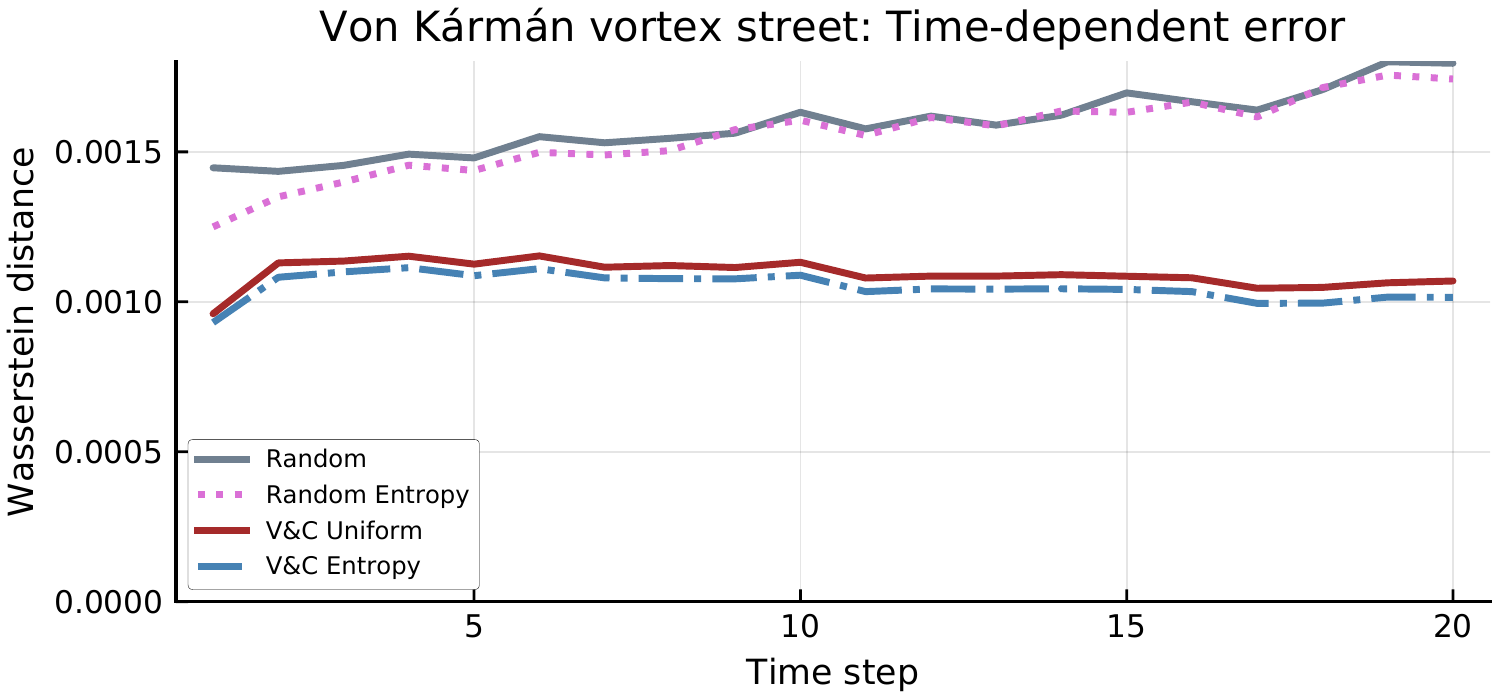}
	\vspace*{-2mm}
	\caption{Comparison of the mean error over all time steps in the von K\'{a}rm\'{a}n vortex street dataset.}
	\label{fig:vk2D:time_error}
\end{figure}

\begin{figure}[t]
	\includegraphics[width=0.49\linewidth]{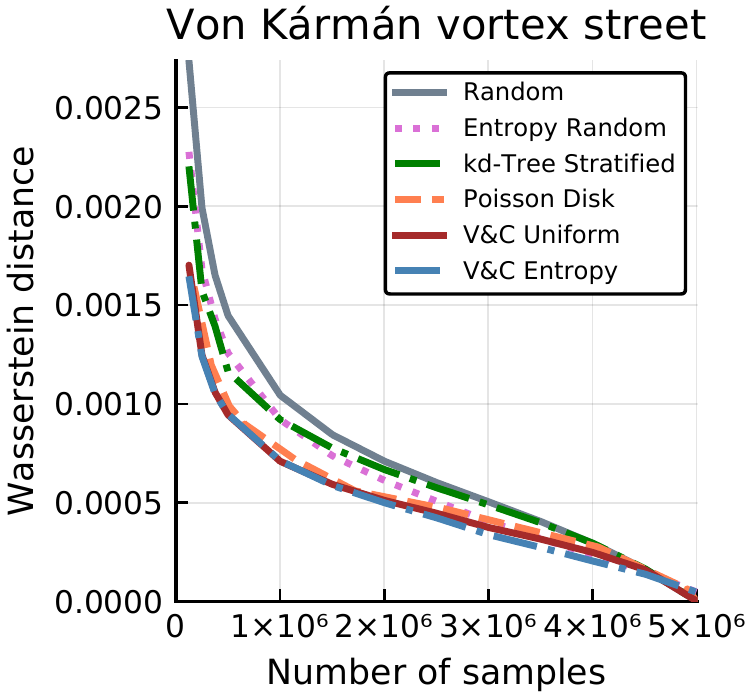}
	\includegraphics[width=0.49\linewidth]{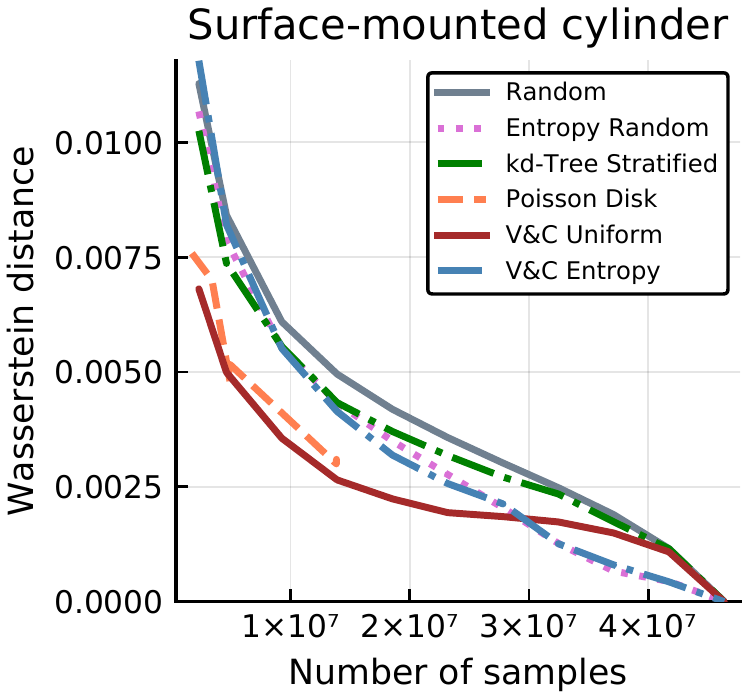}
	\vspace*{-1mm}
	\caption{
		\revi{
			Comparison of the local error measure after sampling a single timestep of the von K\'{a}rm\'{a}n vortex street (left) and the surface-mounted cylinder (right).}}
	\label{fig:vk:error}
\end{figure}

\begin{figure*}
	\subfloat[Random sampling and reconstruction in $t_{10}$]{%
		\includegraphics[width=0.49\linewidth]{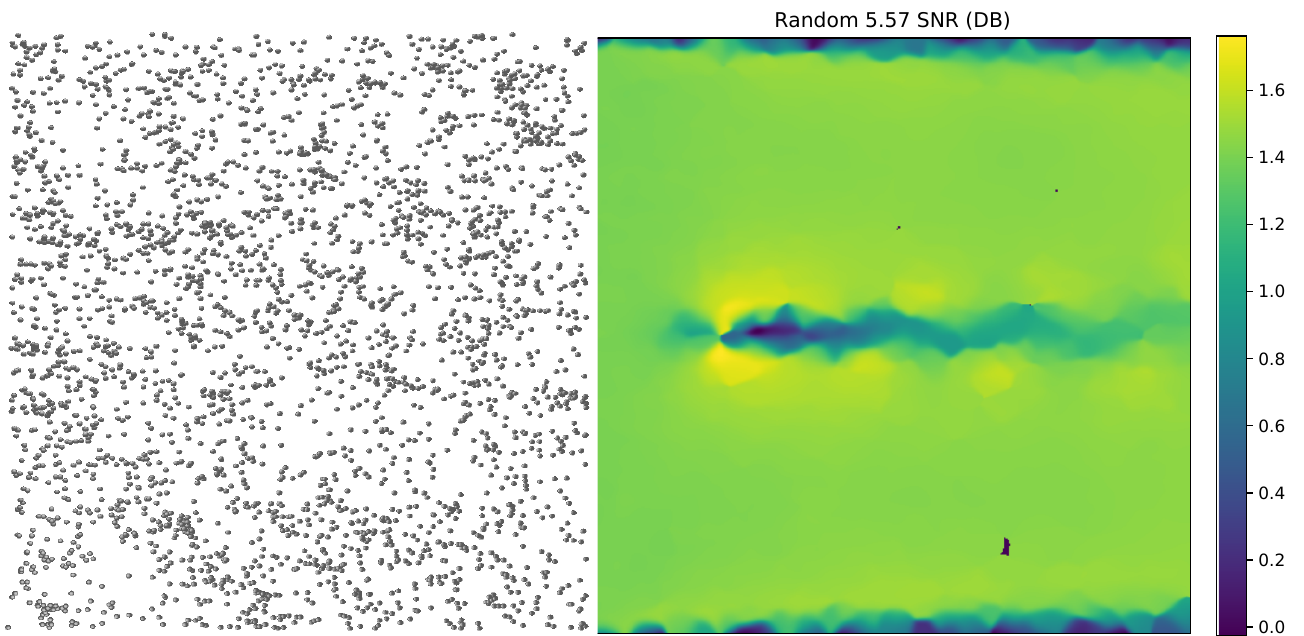}
	}
	\hfill
	\subfloat[Void-and-cluster uniform sampling and reconstruction in $t_{10}$]{%
		\includegraphics[width=0.49\linewidth]{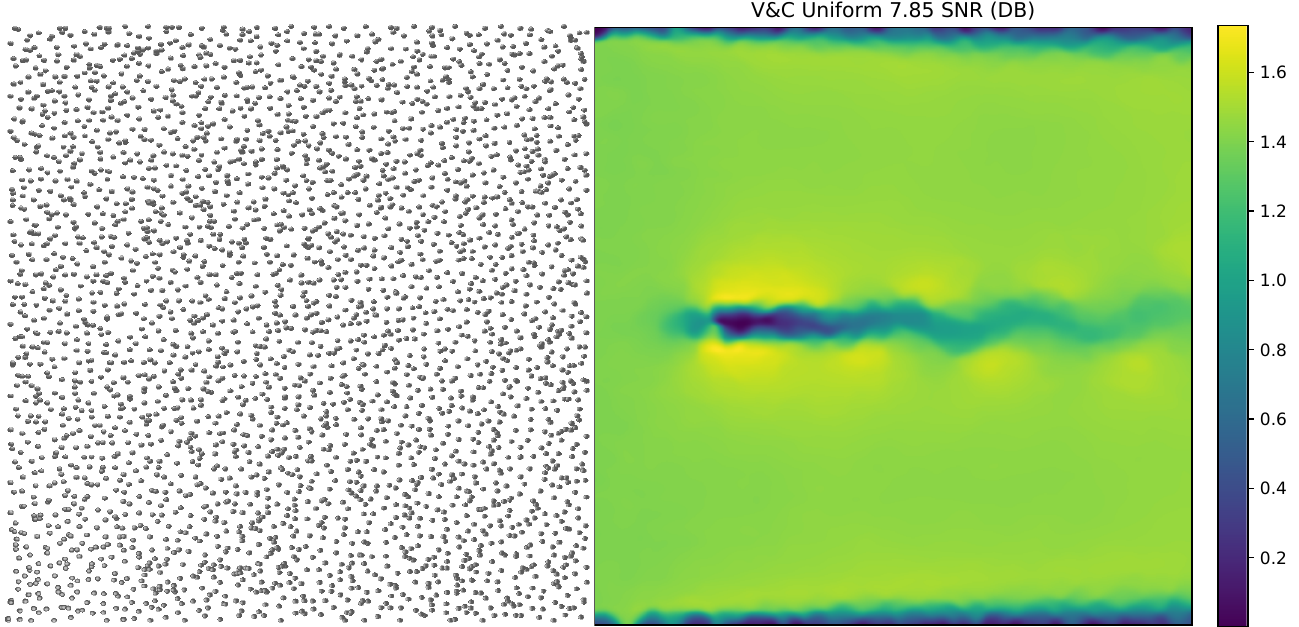}
	}
	\\
	\subfloat[Void-and-cluster entropy sampling and reconstruction in $t_{10}$]{%
		\includegraphics[width=0.49\linewidth]{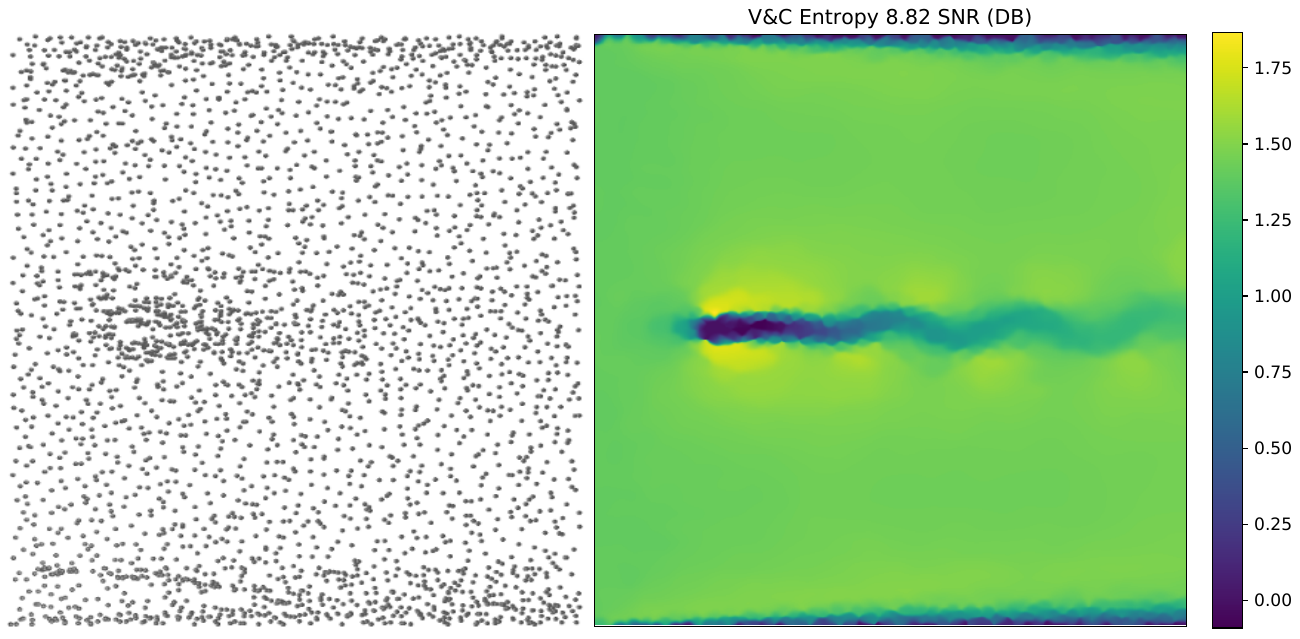}
	}
	\hfill
	\subfloat[Entropy sampling error in $t_{10}$]{%
		\includegraphics[height=4.17cm]{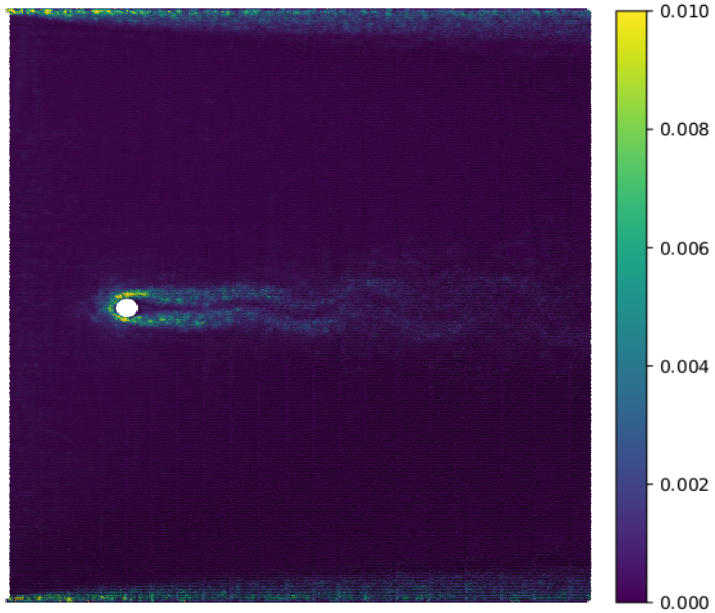}
	}
	\hfill
	\subfloat[Entropy sampling distribution in $t_1$]{%
		\includegraphics[height=4.2cm]{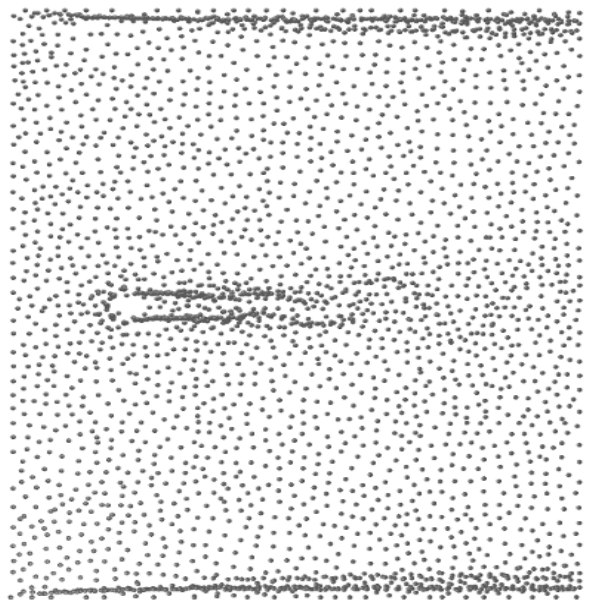}
	}
	\caption{The von K\'{a}rm\'{a}n vortex street after ten time steps using random (a), uniform (b), and entropy (c) void-and-cluster trajectory sampling. The corresponding $u$-velocity fields are shown, which have been created using scattered data interpolation. Our error measure is shown in (d). The entropy void-and-cluster sample distribution in the first time step is illustrated in (e).}
	\label{fig:vk2D:dist}
\end{figure*}

\begin{figure*}
	\subfloat[We visualize the particles using sphere and arrow glyphs with our level-of-detail]{%
		\includegraphics[height=3.6cm]{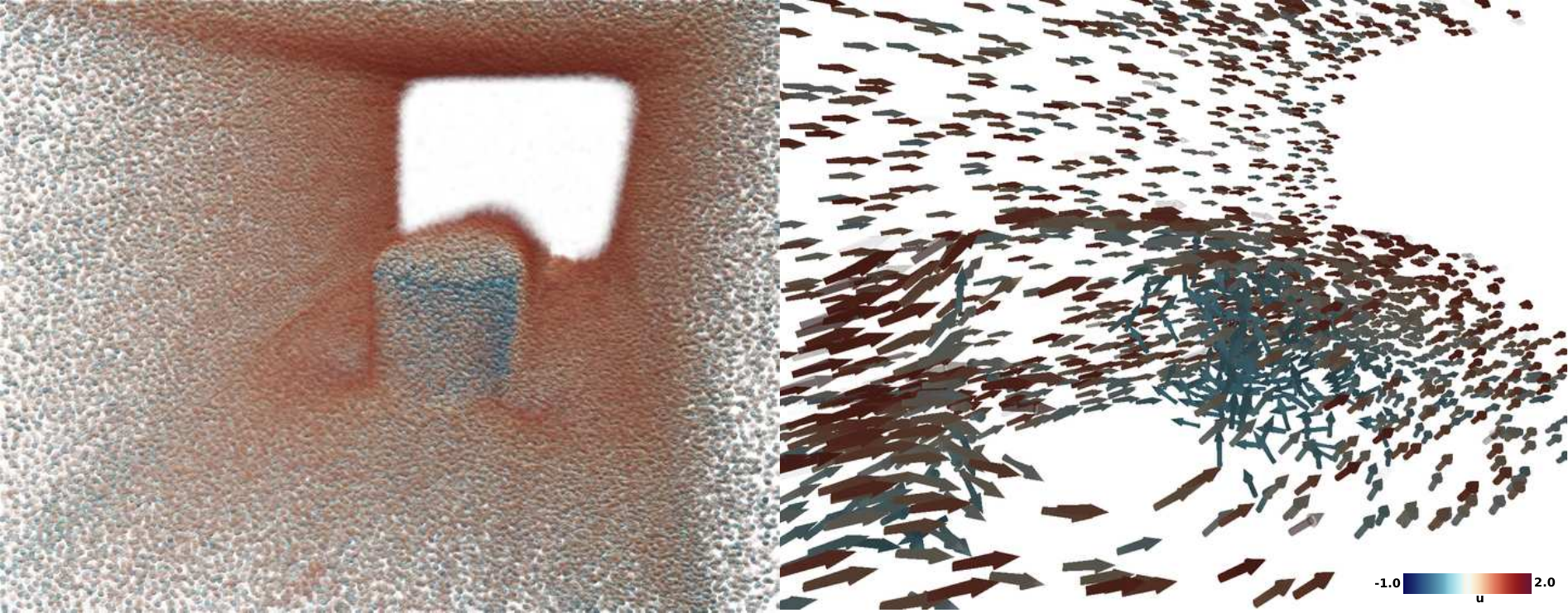}
	}
	\hfill
	\subfloat[Uniform sampling]{%
		\includegraphics[height=3.6cm]{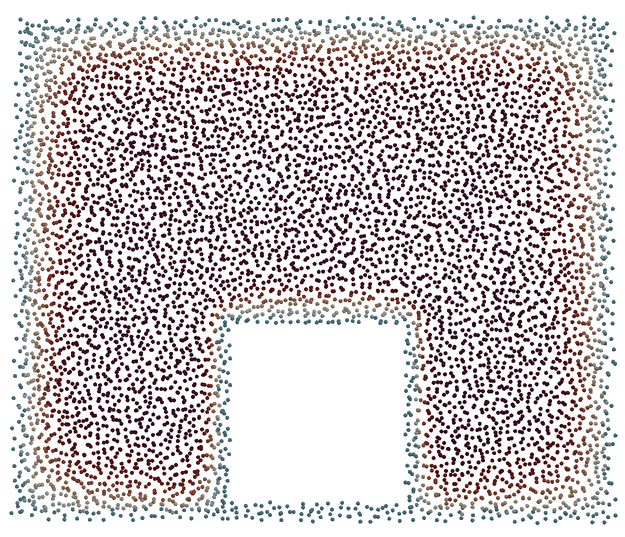}
	}
	\hfill
	\subfloat[Entropy sampling]{%
		\includegraphics[height=3.6cm]{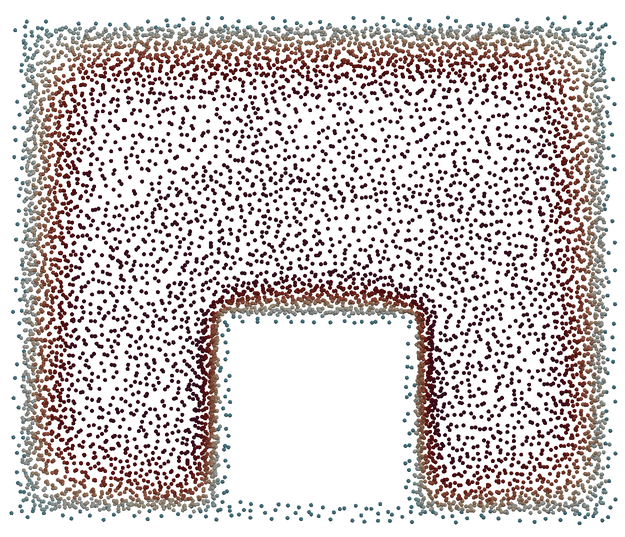}
	}
	\caption{The surface-mounted cylinder, after sampling $466,103$ particles, is shown in (a). We use the continuous level-of-detail, in addition to a transfer function, to further reduce the amount of particles. Slices of the dataset using the uniform (b) and entropy (c) sampling illustrate the difference between the sampling strategies. The entropy strategy samples the less interesting region above the empty cylinder less densely.}
	\label{fig:vk3D}
\end{figure*}

\begin{figure*}
	\subfloat[Histogram from $466,103$ samples]{%
		\includegraphics[height=4.1cm]{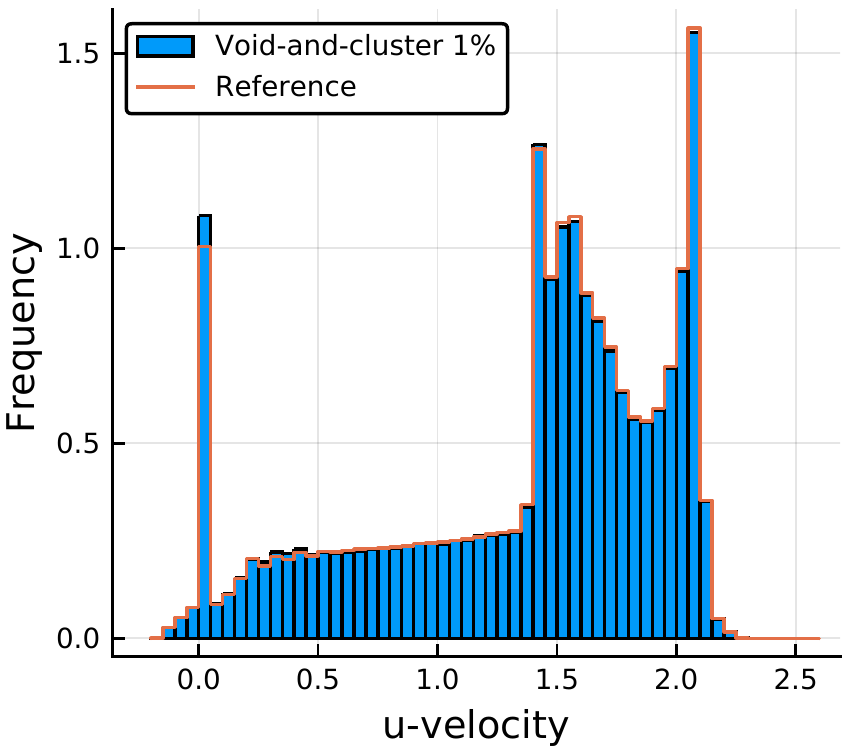}
	}
	\hfill
	\subfloat[Histogram from $10,000$ samples]{%
		\includegraphics[height=4.1cm]{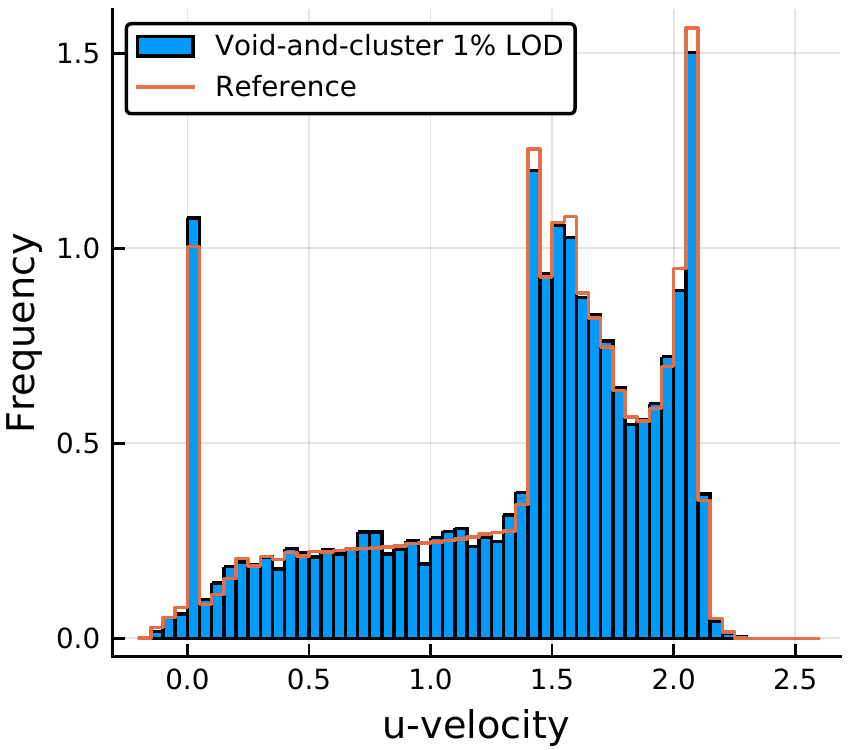}
	}
	\hfill
	\subfloat[Scatter plot from $466,103$ samples]{%
		\includegraphics[height=4.2cm]{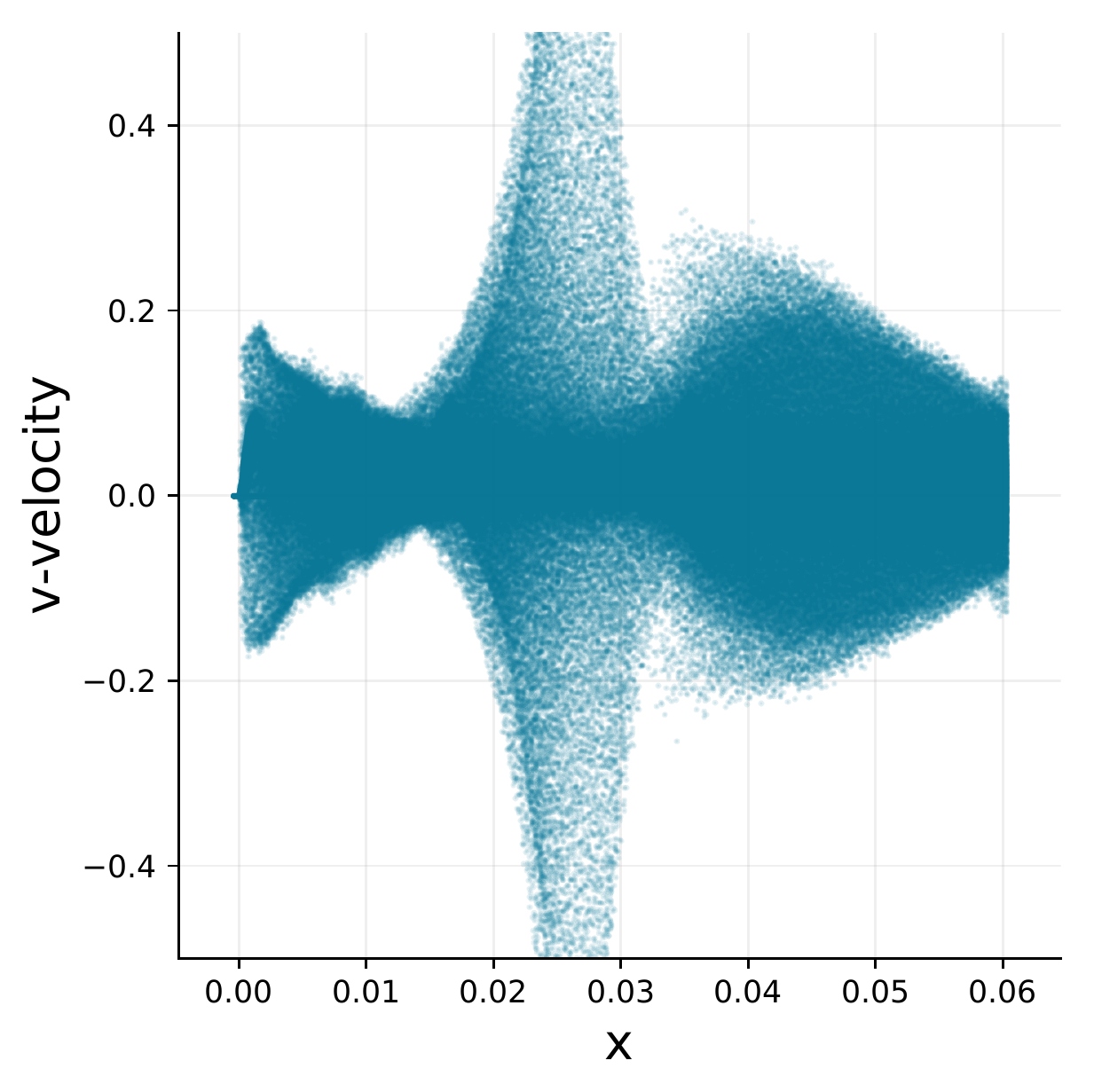}
	}
	\hfill
	\subfloat[Scatter plot from $10,000$ samples]{%
		\includegraphics[height=4.2cm]{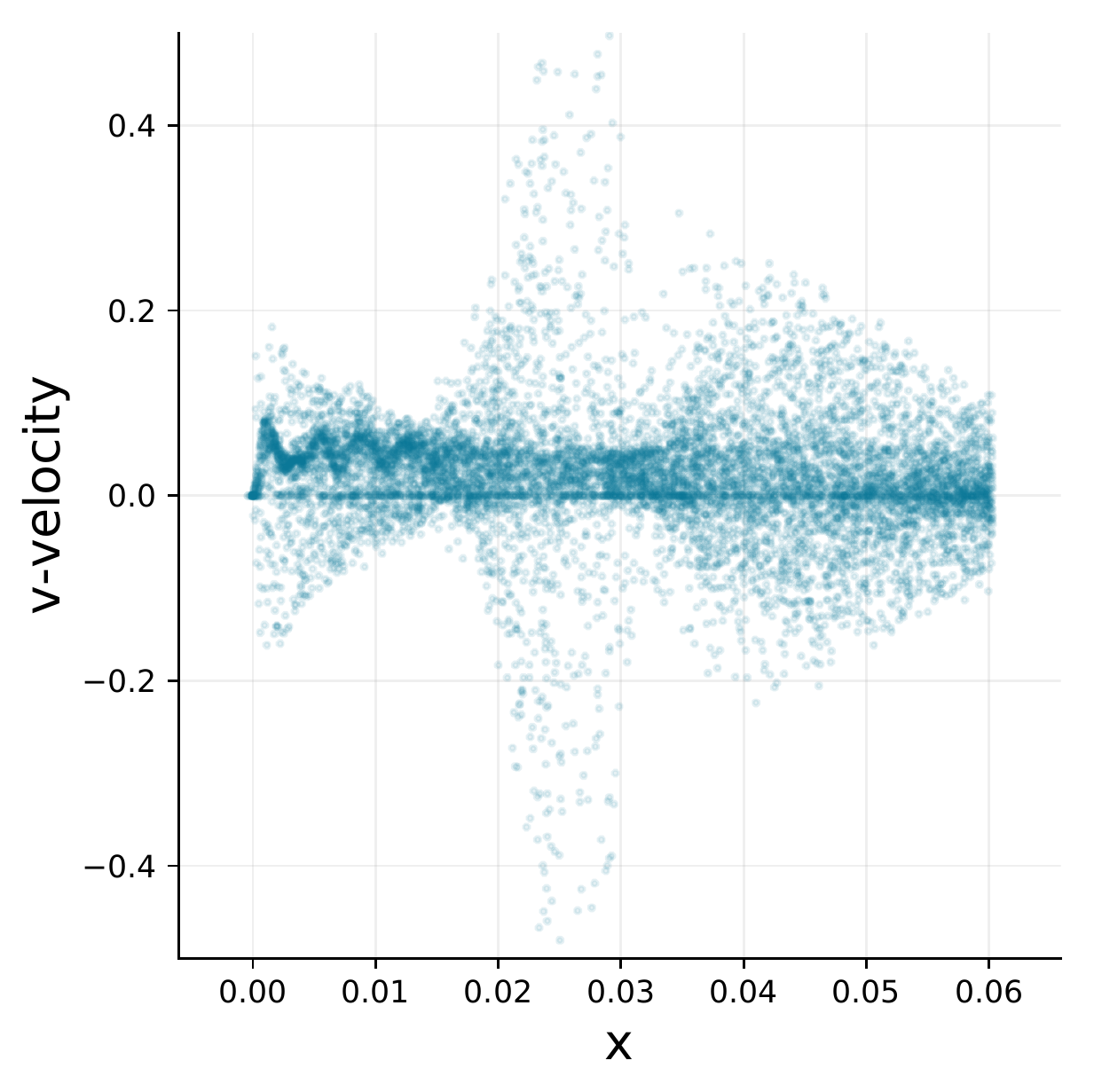}
	}
	\caption{We create a histogram (a) and a scatter plot (c) of the surface-mounted cylinder, after sampling $466,103$ particles using the void-and-cluster entropy strategy. With our level-of-detail, we select a subset of $10.000$ particles and create a histogram (b) and a scatter plot (d).}
	\label{fig:vk3D:dist}
\end{figure*}

The von K\'{a}rm\'{a}n vortex street is a time-dependent SPH dataset that contains about $5$ million particles in each time step. Since the particles enter the domain on the left side and exit on the right, the amount of particles per step changes. A circular boundary in the mid of the domain causes a repeating pattern of swirling vortices, the vortex street.

\revi{
We compare the error measured from the different techniques for sampling the first time step in \autoref{fig:vk:error} (left). Since the original dataset already contains well distributed samples, as a result of the SPH simulation, the Poisson disk sampling can correctly sample the dataset even for large sample counts.
Still, the void-and-cluster techniques consistently lead to the lowest error. The decreased error of stratified kd-tree sampling and entropy random sampling compared to naive random sampling implies that both stratification and entropy-based sampling are beneficial for this dataset.
}

We sample \SI{10}{\percent} of the trajectories in the discrete time interval $[0, 20]$.
%We set the kernel size to $h = \num{1e-4}$, i.e.\ equal to the smoothing length from the simulation.
Since particles frequently enter and exit the domain, we do not explicitly stop trajectories to sample new ones.
In \autoref{fig:vk2D:time_error} we plot the error over time for the different sampling strategies.
The void-and-cluster strategy is not only consistently better, but also stays nearly constant over time.
In contrast, we observe an increase of the error over time for the random sampling techniques. Lastly, the entropy-based sampling strategies show a noticeable improvement for this dataset because they are able to focus more samples on the difficult vortex and boundary regions.

A comparison between random sampling, uniform, and entropy void-and-cluster sampling after $10$ time steps is shown in \autoref{fig:vk2D:dist}. \revi{Although a}ll sampling strategies deteriorate slightly over time, the samples are still well distributed for the uniform void-and-cluster sampling approach. We reconstruct the $u$-velocity field using scattered data interpolation. The results are considerably better for the void-and-cluster approaches. Moreover, the entropy sampling strategy leads to a better reconstruction compared to the uniform void-and-cluster technique. Our error measure of the entropy strategy is shown in (d). Although the entropy strategy already places most samples near the vortex street, the lower \revi{boundary}, and the upper boundary, the error is still high\revi{est} in these regions.

We illustrate the sample distribution for the entropy sampling technique in the first time step in \autoref{fig:vk2D:dist} (e). More samples are placed behind the circular boundary, where vortex shedding occurs, and near the bottom and top of the domain. In these regions, the velocity differs considerably. 
The sampling distribution in the tenth time step has deteriorated considerably for the entropy strategy, but still leads to better results. The entropy strategy thus seems to require shorter trajectories to accurately place samples with respect to the entropy.

\subsection{Surface-Mounted Cylinder}

\begin{figure}
	\centering
	\hfill
	\subfloat[Reference]{%
		\includegraphics[height=2.68cm]{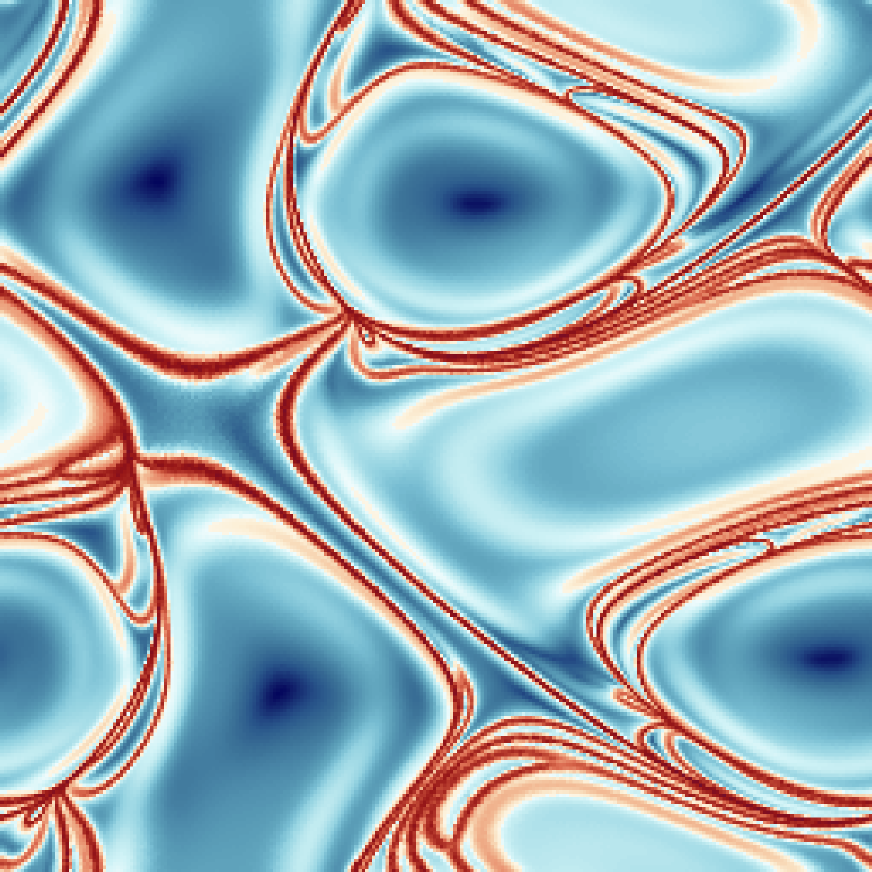}
	}
	\hfill
	\subfloat[Random]{%
		\includegraphics[height=2.68cm]{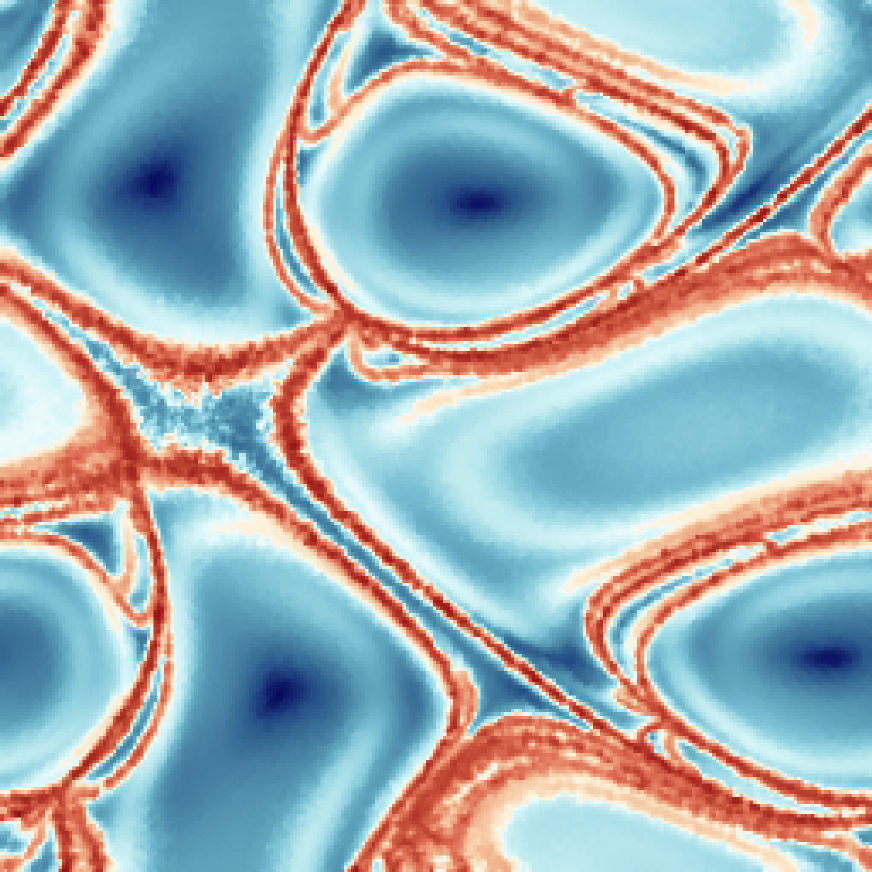}
	}
	\hfill
	\subfloat[Void-and-cluster]{%
		\includegraphics[height=2.68cm]{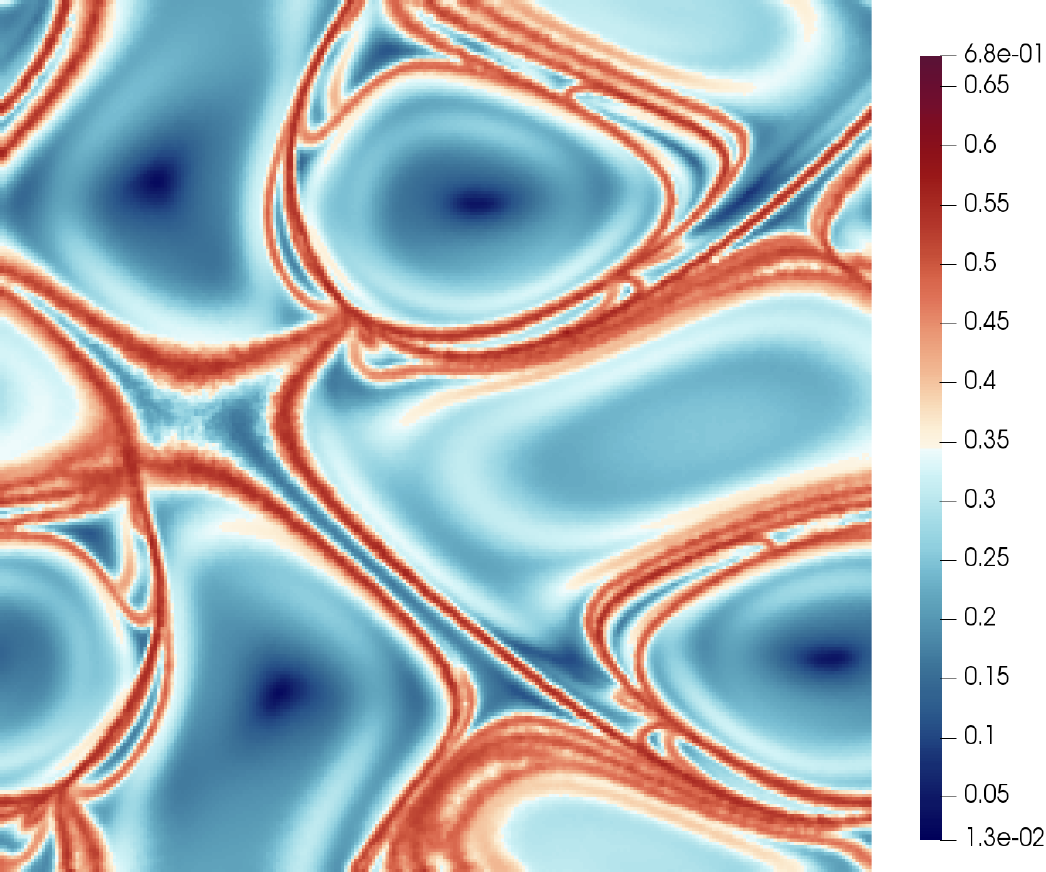}
	}
	\hfill
	\vspace*{-1mm}
	\caption{Slices of the finite-time Lyapunov exponent (FTLE) from the ABC flow. The reference, computed on all trajectories, is shown in (a). We have computed the FTLE after sampling \SI{10}{\percent} of the trajectories by random sampling (b) and using the uniform void-and-cluster (c) technique.}
	\label{fig:ABC_FTLE}
	\vspace*{-1mm}
\end{figure}

This dataset stems from a 3D SPH simulation that simulates the flow around a surface-mounted cylinder~\cite{Sumner2013}. In detail, particles move through a wall-bounded box where a\revi{n empty} cylinder is placed on the bottom. The dataset contains about $46$ million particles in each time step, each of which has a position, velocity, and pressure and either belongs to the static domain boundary or the simulated fluid.
\revi{
In \autoref{fig:vk:error} (right), we compare the error of the different sampling techniques.
Most notably, the entropy-based techniques show a larger error for smaller sample counts.
%Presumably, this is due to large amount of non-uniform sampling that leads to a higher error in any value dimensions.
}

We sample \SI{1}{\percent} of the dataset and visualize the particles as sphere and arrow glyphs in \autoref{fig:vk3D} (a). We map $u$-velocity to color, i.e.\ velocity in the principal flow direction. Since the sampled subset still contains a large amount of particles, we make use of the continuous level-of-detail in addition to a transfer function, which maps fast particles to transparent, to further reduce the amount of visual clutter. The vortex shedding in the wake of the cylinder thus becomes visible.
Furthermore, we illustrate the difference between uniform and entropy void-and-cluster sampling in \autoref{fig:vk3D} (b) and (c). The entropy strategy samples the regions near the wall and close to the cylinder more densely \revi{ due to fluctuating velocities, but also due to the interface between fluid and boundary particles that leads to a high entropy}. \revirm{This behavior is desirable since }\revi{In contrast,} the region\revi{s} above \revi{and next to} the cylinder contain\revirm{s} a large amount of particles that move unobstructed through the domain.

In \autoref{fig:vk3D:dist}, we create a histogram of $u$-velocity (a) and a scatter plot of $x$ and $v$-velocity (c) of the dataset sampled with the entropy void-and-cluster technique. We use our level-of-detail mechanism to create a subset of $10,000$ particles and compute similar plots in (b) and (d). The histograms in (a) and (b) are similar, even though we reduce the amount of samples considerably. In the scatter plot (d), the amount of clutter is significantly reduced. The periodic changes in $v$-velocity, caused by the swirling vortices in the wake of the cylinder, then become visible.
Lastly, the ordering of samples allows us to optimize loading times and latency. In particular, when opening a new file or time step, we initially load only a small subset and asynchronously continue to load more samples to reduce the latency. Especially for larger datasets, we found working with a small subset of the data to be preferable due to the fast and less cluttered visualizations.

\subsection{The ABC Flow}

\begin{figure}[t]
	\includegraphics[width=\linewidth]{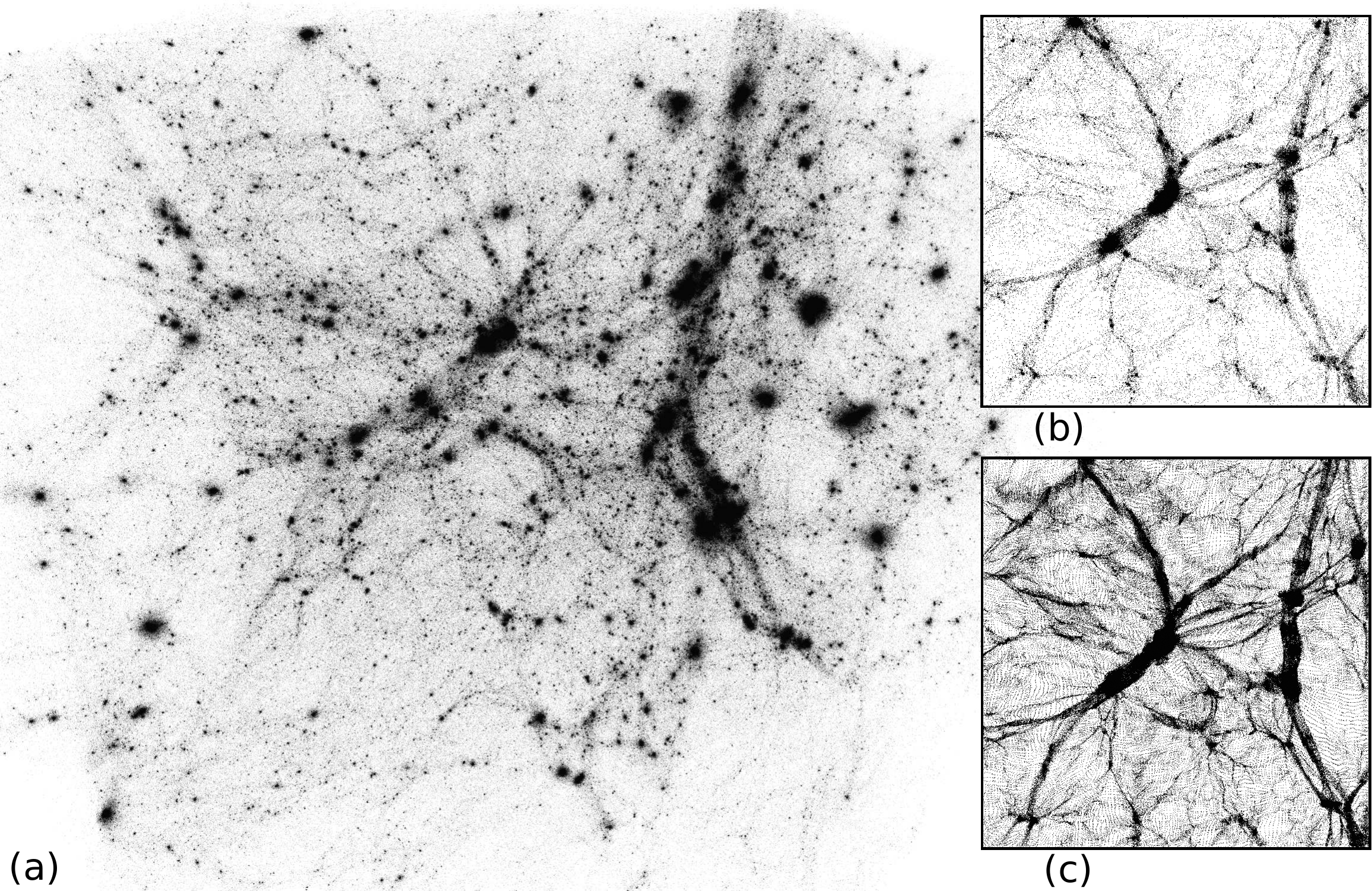}%
	\vspace*{-1mm}
	\caption{\revi{
			The Dark Sky dataset reduced to \SI{5}{\percent} using the uniform void-and-cluster technique (a). A slice of the dataset is shown in (b), with the corresponding slice from the original dataset in (c).}}
	\label{fig:darksky}
%	\vspace*{-1mm}
\end{figure}

\begin{figure*}
	\includegraphics[width=0.24\linewidth]{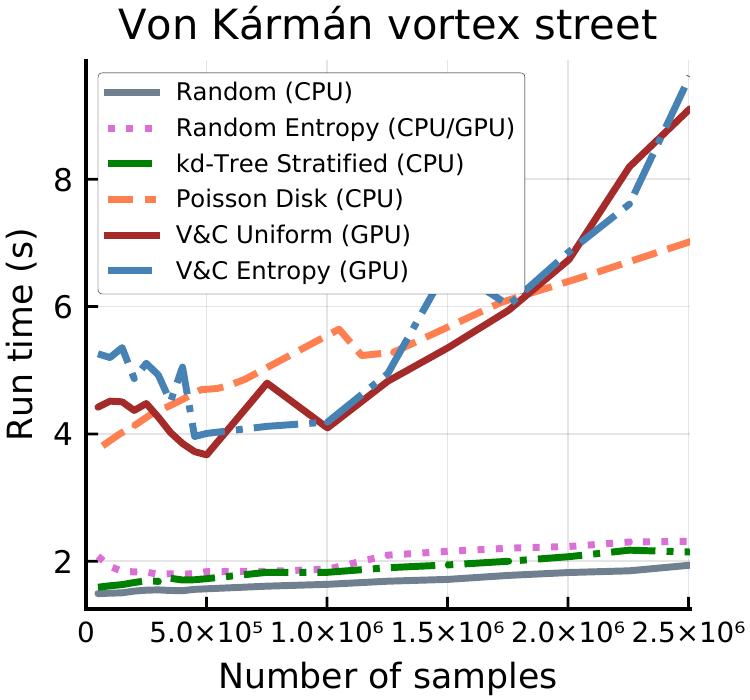}
	\includegraphics[width=0.24\linewidth]{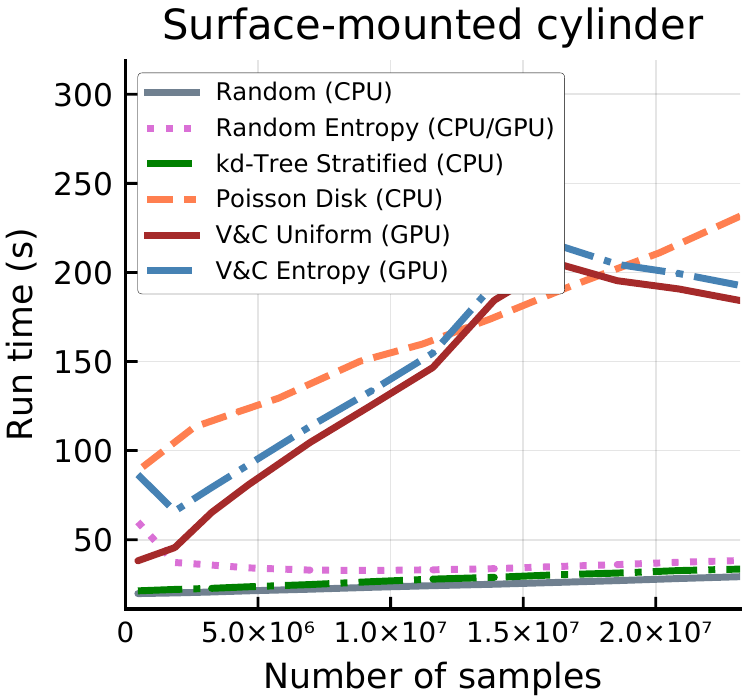}
	\includegraphics[width=0.24\linewidth]{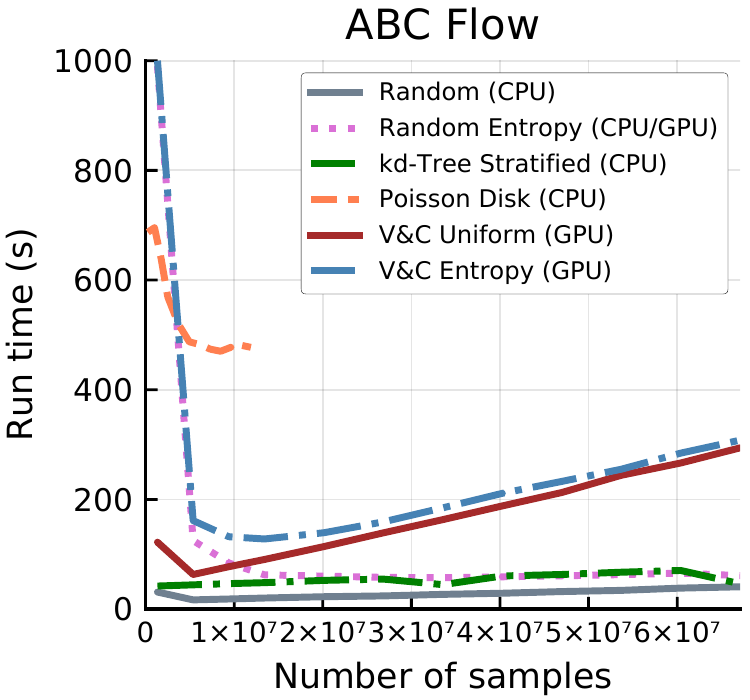}
	\includegraphics[width=0.24\linewidth]{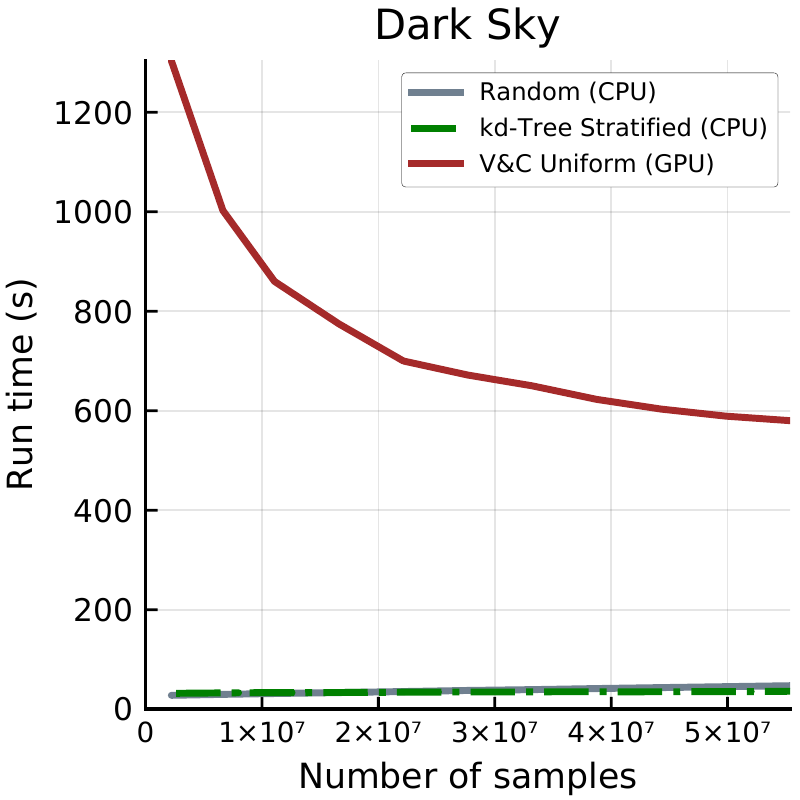}
	\caption{The sampling performance on the von K\'{a}rm\'{a}n vortex street, the surface-mounted cylinder, the ABC flow, and the Dark Sky dataset.}
	\label{fig:runtime}
\end{figure*}

\begin{figure}
	\includegraphics[width=0.5\linewidth]{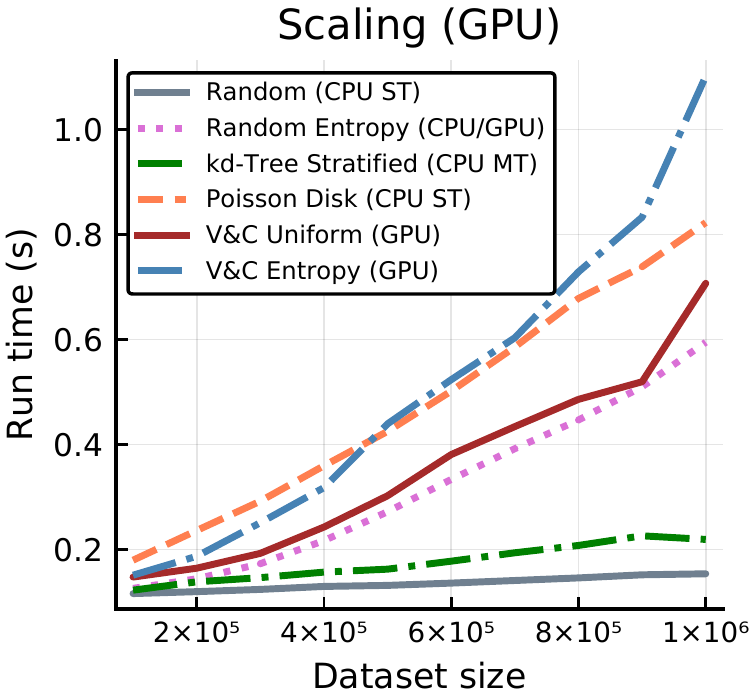}%
	\includegraphics[width=0.5\linewidth]{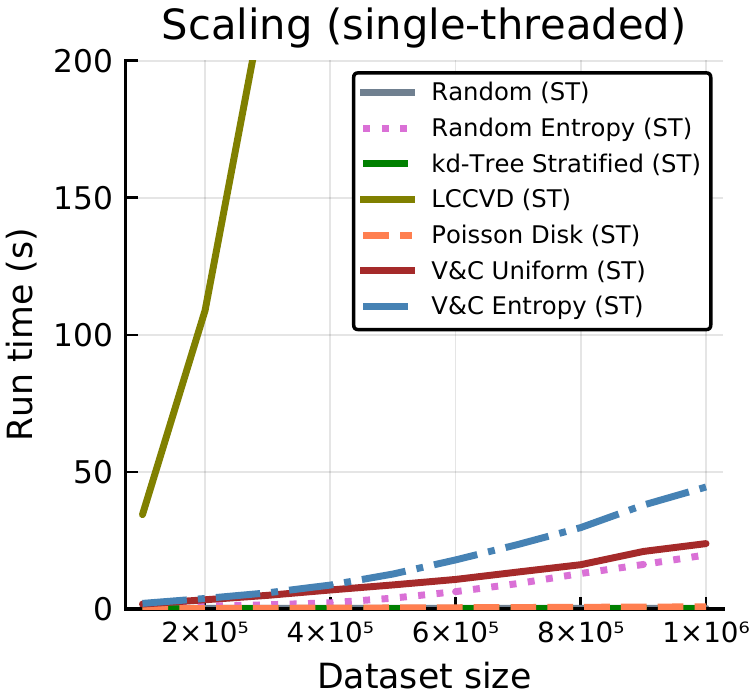}%
	\caption{We evaluate the scalability of our algorithm using differently sized  $\sinc$ datasets, whilst always sampling \SI{10}{\percent}.}
	\label{fig:scaling}
\end{figure}

The Arnold-Beltrami-Childress (ABC) flow is a three-dimensional, steady velocity field:
\begin{align*}
	\dot{x} = A \sin z + C \cos y \\
	\dot{y} = B \sin x + A \cos z \\
	\dot{z} = C \sin y + B \cos x.
\end{align*}
We set $A = \sqrt{3}$, $B = \sqrt{2}$, $C = 1$. We represent the flow in the Lagrangian basis with $134,217,728$ trajectories that start in the spatial domain $[0, 2\pi]^3$ and are integrated using a 4th-order Runge-Kutta scheme over the time interval $[0, 10]$. We then sample \SI{10}{\percent} of the trajectories\revi{, without stopping and starting new trajectories}.

In \autoref{fig:teaser} (g) and (h), $50,000$ and $500$ trajectories are shown as illuminated pathlines using the continuous level-of-detail that is implicitly given by the rank of our sampling strategy. Since we reorder the samples by their rank, we only have to load the first samples for visualization and can load additional samples progressively.

After random sampling and uniform void-and-cluster sampling, we have computed the (forward) finite-time Lyapunov exponent (FTLE). This quantity measures how neighboring trajectories separate over time and is used to visualize time-dependent flow behavior. A slice of the FTLE is shown in \autoref{fig:ABC_FTLE}. 
Computing the FTLE after sampling with the uniform void-and-cluster strategy yields better results compared to random sampling, even though the same number of samples have been taken. 

\revi{
\subsection{Dark Sky}

The Dark Sky simulations are a series of cosmological N-body simulations of the evolution of the large-scale universe~\cite{Skillman2014}. 
We study a subset that consists of 111 million particles with a position, velocity, and unique identifier. \autoref{fig:darksky} shows a visualization of the dataset reduced to \SI{5}{\percent}.
Since the simulations investigate the clustering of particles into galaxies, filaments, and the emergence of cosmic voids, the spatial distribution of the particles is strongly non-uniform. Consequently, a sampled subset of the data should preserve this distribution of cosmological mass. This is not possible using Poisson disk sampling or entropy-based adaptive sampling.
In contrast, our uniform void-and-cluster technique optimizes the blue noise property with respect to the spatial density of particles in the dataset. The spatial distribution is thus preserved, whilst the samples are optimally stratified. 
}

\subsection{Performance and Scalability}

To assess the performance of our algorithm, we compare the run time of different sampling strategies. For the measurements, we use an Intel Core i7-6700 and an Nvidia Quadro RTX 8000. We enable GPU acceleration where possible.

Measurements for all of our datasets are shown in \autoref{fig:runtime}. We were not able to measure our implementation of LCCVD for larger datasets since it is computationally demanding and we did not parallelize it.
\revi{Although Poisson disk sampling is a linear time algorithm, it is inherently sequential and leads to long run times for large data sizes.}
Random and stratified sampling are fast even though no GPU acceleration is used. In comparison, the void-and-cluster techniques are slower, but considering the data sizes we argue that the performance is acceptable. For example, we take $1,342,177$ samples out of $134$ million from the ABC flow dataset in $68$ seconds.
\revi{Due to the non-uniform input data of the dark sky simulation, a large kernel support is required that leads to a significantly increased run time. The use of adaptive kernel sizes or better suited data structures could potentially improve the efficiency of the neighborhood search for non-uniformly distributed datasets.}

The uniform and entropy-based sampling strategies perform similar. \revi{However, f}or small sampling percentages\revirm{, however,} the entropy computation is noticeably slower since the computation depends on the kernel size $\kernelSizeSamples$, which increases for smaller sampling percentages\revi{, and scales with the input data size. This is especially visible in the ABC flow where the run-time of the entropy computation increases dramatically for smaller sampling percentages due to the neighborhood lookup limiting the GPU efficiency.}
In general, the run-time increases when a larger number of samples is taken, which indicates that the void filling step is the bottleneck. 

Lastly, to measure the time complexity and scalability of the algorithm, we measure GPU and single-threaded CPU performance with differently sized $\sinc$ datasets. The measurements are shown in \autoref{fig:scaling}.
Although our GPU implementation is competitive with random and stratified sampling, the single-threaded CPU implementation is considerably slower. This highlights the benefits of parallelization and GPU acceleration for our algorithm. Compared to LCCVD\revirm{, however,} our single-threaded implementation achieves an enormous speed-up. Although not shown in the plot, LCCVD took more than $2,500$ seconds to sample a dataset of size $10^6$.

\revi{
\section{Future Work: Multi-Node Parallelism}
An important use case that has not been addressed so far is the parallel implementation for distributed memory systems.
To scale the sampling technique to multiple nodes on a compute cluster, the spatial domain could be subdivided into uniform tiles. Each compute node then performs void-and-cluster sampling of one tile. Although this will produce tiles that are well distributed according to the blue noise property, the samples in the border region between two or more tiles are not necessarily well separated. If this is not acceptable, then the cluster-and-void optimization step can be applied again to the whole dataset. However, this would have to be performed on a single node which might not be possible, for example due to memory constraints. The extension of the proposed algorithm for multi-node parallelism is thus still an open problem.
}

\section{Conclusion}

We present a novel approach for using statistical sampling to reduce large scattered datasets for visualization and analysis. In particular, our void-and-cluster technique optimizes sample distributions with respect to the blue noise property and thus produces samples that evenly cover the spatio-temporal domain. Our technique significantly improves the accuracy of operations such as scattered-data interpolation or the computation of the FTLE\@. In combination with the level-of-detail given by our technique, we are able to generate interactive and clutter-free visualizations of large datasets.
Lastly, we introduce an error measure to quantify the error of a subsampled dataset, which takes both spatial and value domain into account. Our results show a clear correlation between the error measure and the quality of derived quantities such as scattered data interpolation.

%% if specified like this the section will be committed in review mode
\acknowledgments{
The von K\'{a}rm\'{a}n vortex street and the surface-mounted cylinder datasets are courtesy of Thilo Dauch and Rainer Koch from the Institute of Thermal Turbomachinery at the Karlsruhe Institute of Technology.
}

\bibliographystyle{abbrv-doi}

\bibliography{../data_sampling}

\begin{thebibliography}{10}

\bibitem{Agranovsky2014}
A.~Agranovsky, D.~Camp, C.~Garth, E.~W. Bethel, K.~I. Joy, and H.~Childs.
\newblock Improved post hoc flow analysis via lagrangian representations.
\newblock In {\em IEEE Symposium on Large Data Analysis and Visualization}, pp.
  67--75, 2014. doi: {{%
10\hspace{.1pt}\discretionary{.}{%
}{.}\hspace{.4pt}1109\discretionary{/}{%
}{/}LDAV\hspace{.1pt}\discretionary{.}{%
}{.}\hspace{.4pt}2014\hspace{.1pt}\discretionary{.}{%
}{.}\hspace{.4pt}7013206}}


\bibitem{Balzer2009}
M.~Balzer, T.~Schl\"{o}mer, and O.~Deussen.
\newblock Capacity-constrained point distributions: A variant of lloyd's
  method.
\newblock {\em ACM Transactions on Graphics}, 28(3):86:1--86:8, 2009. doi: {{%
10\hspace{.1pt}\discretionary{.}{%
}{.}\hspace{.4pt}1145\discretionary{/}{%
}{/}1531326\hspace{.1pt}\discretionary{.}{%
}{.}\hspace{.4pt}1531392}}


\bibitem{Biswas2018}
A.~Biswas, S.~Dutta, J.~Pulido, and J.~Ahrens.
\newblock In situ data-driven adaptive sampling for large-scale simulation data
  summarization.
\newblock In {\em Proceedings of the Workshop on In Situ Infrastructures for
  Enabling Extreme-Scale Analysis and Visualization}, pp. 13--18. ACM, 2018.
  doi: {{%
10\hspace{.1pt}\discretionary{.}{%
}{.}\hspace{.4pt}1145\discretionary{/}{%
}{/}3281464\hspace{.1pt}\discretionary{.}{%
}{.}\hspace{.4pt}3281467}}


\bibitem{Bridson2007}
R.~Bridson.
\newblock Fast poisson disk sampling in arbitrary dimensions.
\newblock In {\em ACM SIGGRAPH 2007 Sketches}, SIGGRAPH 2007, p.~22. ACM, 2007.
  doi: {{%
10\hspace{.1pt}\discretionary{.}{%
}{.}\hspace{.4pt}1145\discretionary{/}{%
}{/}1278780\hspace{.1pt}\discretionary{.}{%
}{.}\hspace{.4pt}1278807}}


\bibitem{Dix2002}
A.~Dix and G.~Ellis.
\newblock By chance enhancing interaction with large data sets through
  statistical sampling.
\newblock In {\em Proceedings of the Working Conference on Advanced Visual
  Interfaces}, pp. 167--176, 2002. doi: {{%
10\hspace{.1pt}\discretionary{.}{%
}{.}\hspace{.4pt}1145\discretionary{/}{%
}{/}1556262\hspace{.1pt}\discretionary{.}{%
}{.}\hspace{.4pt}1556289}}


\bibitem{Dutta2017a}
S.~Dutta, C.~M. Chen, G.~Heinlein, H.~W. Shen, and J.~P. Chen.
\newblock In situ distribution guided analysis and visualization of transonic
  jet engine simulations.
\newblock {\em IEEE Transactions on Visualization and Computer Graphics},
  23(1):811--820, 2017. doi: {{%
10\hspace{.1pt}\discretionary{.}{%
}{.}\hspace{.4pt}1109\discretionary{/}{%
}{/}TVCG\hspace{.1pt}\discretionary{.}{%
}{.}\hspace{.4pt}2016\hspace{.1pt}\discretionary{.}{%
}{.}\hspace{.4pt}2598604}}


\bibitem{Dutta2017}
S.~Dutta, J.~Woodring, H.~W. Shen, J.~P. Chen, and J.~Ahrens.
\newblock Homogeneity guided probabilistic data summaries for analysis and
  visualization of large-scale data sets.
\newblock In {\em IEEE Pacific Visualization Symposium}, pp. 111--120, 2017.
  doi: {{%
10\hspace{.1pt}\discretionary{.}{%
}{.}\hspace{.4pt}1109\discretionary{/}{%
}{/}PACIFICVIS\hspace{.1pt}\discretionary{.}{%
}{.}\hspace{.4pt}2017\hspace{.1pt}\discretionary{.}{%
}{.}\hspace{.4pt}8031585}}


\bibitem{Fraedrich2010}
R.~{Fraedrich}, S.~{Auer}, and R.~{Westermann}.
\newblock Efficient high-quality volume rendering of sph data.
\newblock {\em IEEE Transactions on Visualization and Computer Graphics},
  16(6):1533--1540, 2010. doi: {{%
10\hspace{.1pt}\discretionary{.}{%
}{.}\hspace{.4pt}1109\discretionary{/}{%
}{/}TVCG\hspace{.1pt}\discretionary{.}{%
}{.}\hspace{.4pt}2010\hspace{.1pt}\discretionary{.}{%
}{.}\hspace{.4pt}148}}


\bibitem{Frey2011}
S.~Frey, T.~Schl\"{o}mer, S.~Grottel, C.~Dachsbacher, O.~Deussen, and T.~Ertl.
\newblock Loose capacity-constrained representatives for the qualitative visual
  analysis in molecular dynamics.
\newblock In {\em IEEE Pacific Visualization Symposium}, pp. 51--58, 2011. doi:
  {{%
10\hspace{.1pt}\discretionary{.}{%
}{.}\hspace{.4pt}1109\discretionary{/}{%
}{/}PACIFICVIS\hspace{.1pt}\discretionary{.}{%
}{.}\hspace{.4pt}2011\hspace{.1pt}\discretionary{.}{%
}{.}\hspace{.4pt}5742372}}


\bibitem{Grottel2014}
S.~Grottel, M.~Krone, C.~M\"{u}ller, G.~Reina, and T.~Ertl.
\newblock Megamol -- a prototyping framework for particle-based visualization.
\newblock {\em IEEE Transactions on Visualization and Computer Graphics},
  21(2):201--214, 2015. doi: {{%
10\hspace{.1pt}\discretionary{.}{%
}{.}\hspace{.4pt}1109\discretionary{/}{%
}{/}TVCG\hspace{.1pt}\discretionary{.}{%
}{.}\hspace{.4pt}2014\hspace{.1pt}\discretionary{.}{%
}{.}\hspace{.4pt}2350479}}


\bibitem{Hazarika2019}
S.~Hazarika, S.~Dutta, H.~Shen, and J.~Chen.
\newblock Codda: A flexible copula-based distribution driven analysis framework
  for large-scale multivariate data.
\newblock {\em IEEE Transactions on Visualization and Computer Graphics},
  25(1):1214--1224, 2019. doi: {{%
10\hspace{.1pt}\discretionary{.}{%
}{.}\hspace{.4pt}1109\discretionary{/}{%
}{/}TVCG\hspace{.1pt}\discretionary{.}{%
}{.}\hspace{.4pt}2018\hspace{.1pt}\discretionary{.}{%
}{.}\hspace{.4pt}2864801}}


\bibitem{Hlawatsch2011}
M.~Hlawatsch, F.~Sadlo, and D.~Weiskopf.
\newblock Hierarchical line integration.
\newblock {\em IEEE Transactions on Visualization and Computer Graphics},
  17(8):1148--1163, 2011. doi: {{%
10\hspace{.1pt}\discretionary{.}{%
}{.}\hspace{.4pt}1109\discretionary{/}{%
}{/}TVCG\hspace{.1pt}\discretionary{.}{%
}{.}\hspace{.4pt}2010\hspace{.1pt}\discretionary{.}{%
}{.}\hspace{.4pt}227}}


\bibitem{Li2018}
S.~Li, N.~Marsaglia, C.~Garth, J.~Woodring, J.~Clyne, and H.~Childs.
\newblock Data reduction techniques for simulation, visualization and data
  analysis.
\newblock {\em Computer Graphics Forum}, 37(6):422--447, 2018. doi: {{%
10\hspace{.1pt}\discretionary{.}{%
}{.}\hspace{.4pt}1111\discretionary{/}{%
}{/}cgf\hspace{.1pt}\discretionary{.}{%
}{.}\hspace{.4pt}13336}}


\bibitem{Monaghan1982}
J.~J. Monaghan.
\newblock Why particle methods work.
\newblock {\em SIAM Journal on Scientific and Statistical Computing},
  3(4):422--433, 1982. doi: {{%
10\hspace{.1pt}\discretionary{.}{%
}{.}\hspace{.4pt}1137\discretionary{/}{%
}{/}0903027}}


\bibitem{Monaghan1992}
J.~J. Monaghan.
\newblock Smoothed particle hydrodynamics.
\newblock {\em Annual review of astronomy and astrophysics}, 30(1):543--574,
  1992. doi: {{%
10\hspace{.1pt}\discretionary{.}{%
}{.}\hspace{.4pt}1146\discretionary{/}{%
}{/}annurev\hspace{.1pt}\discretionary{.}{%
}{.}\hspace{.4pt}aa\hspace{.1pt}\discretionary{.}{%
}{.}\hspace{.4pt}30\hspace{.1pt}\discretionary{.}{%
}{.}\hspace{.4pt}090192\hspace{.1pt}\discretionary{.}{%
}{.}\hspace{.4pt}002551}}


\bibitem{Reichl2013}
F.~{Reichl}, M.~{Treib}, and R.~{Westermann}.
\newblock Visualization of big sph simulations via compressed octree grids.
\newblock In {\em IEEE International Conference on Big Data}, pp. 71--78, 2013.
  doi: {{%
10\hspace{.1pt}\discretionary{.}{%
}{.}\hspace{.4pt}1109\discretionary{/}{%
}{/}BigData\hspace{.1pt}\discretionary{.}{%
}{.}\hspace{.4pt}2013\hspace{.1pt}\discretionary{.}{%
}{.}\hspace{.4pt}6691717}}


\bibitem{Reinhardt2017}
S.~{Reinhardt}, M.~{Huber}, O.~{Dumitrescu}, M.~{Krone}, B.~{Eberhardt}, and
  D.~{Weiskopf}.
\newblock Visual debugging of sph simulations.
\newblock In {\em 21st International Conference Information Visualisation}, pp.
  117--126, 2017. doi: {{%
10\hspace{.1pt}\discretionary{.}{%
}{.}\hspace{.4pt}1109\discretionary{/}{%
}{/}iV\hspace{.1pt}\discretionary{.}{%
}{.}\hspace{.4pt}2017\hspace{.1pt}\discretionary{.}{%
}{.}\hspace{.4pt}20}}


\bibitem{Sauer2017}
F.~{Sauer}, J.~{Xie}, and K.~{Ma}.
\newblock A combined eulerian-lagrangian data representation for large-scale
  applications.
\newblock {\em IEEE Transactions on Visualization and Computer Graphics},
  23(10):2248--2261, 2017. doi: {{%
10\hspace{.1pt}\discretionary{.}{%
}{.}\hspace{.4pt}1109\discretionary{/}{%
}{/}TVCG\hspace{.1pt}\discretionary{.}{%
}{.}\hspace{.4pt}2016\hspace{.1pt}\discretionary{.}{%
}{.}\hspace{.4pt}2620975}}


\bibitem{Sicat2014}
R.~Sicat, J.~Kr{\"u}ger, T.~M{\"o}ller, and M.~Hadwiger.
\newblock Sparse pdf volumes for consistent multi-resolution volume rendering.
\newblock {\em IEEE Transactions on Visualization and Computer Graphics},
  20(12):2417--2426, 2014. doi: {{%
10\hspace{.1pt}\discretionary{.}{%
}{.}\hspace{.4pt}1109\discretionary{/}{%
}{/}TVCG\hspace{.1pt}\discretionary{.}{%
}{.}\hspace{.4pt}2014\hspace{.1pt}\discretionary{.}{%
}{.}\hspace{.4pt}2346324}}


\bibitem{Skillman2014}
S.~W. Skillman, M.~S. Warren, M.~J. Turk, R.~H. Wechsler, D.~E. Holz, and P.~M.
  Sutter.
\newblock Dark sky simulations: Early data release, 2014.
\newblock arXiv:1407.2600.

\bibitem{Su2014}
Y.~Su, G.~Agrawal, J.~Woodring, K.~Myers, J.~Wendelberger, and J.~Ahrens.
\newblock Effective and efficient data sampling using bitmap indices.
\newblock {\em Cluster Computing}, 17(4):1081--1100, 2014. doi: {{%
10\hspace{.1pt}\discretionary{.}{%
}{.}\hspace{.4pt}1007\discretionary{/}{%
}{/}s10586\discretionary{%
}{-}{-}014\discretionary{%
}{-}{-}0360\discretionary{%
}{-}{-}5}}


\bibitem{Sumner2013}
D.~Sumner.
\newblock Flow above the free end of a surface-mounted finite-height circular
  cylinder: A review.
\newblock {\em Journal of Fluids and Structures}, 43:41 -- 63, 2013. doi: {{%
10\hspace{.1pt}\discretionary{.}{%
}{.}\hspace{.4pt}1016\discretionary{/}{%
}{/}j\hspace{.1pt}\discretionary{.}{%
}{.}\hspace{.4pt}jfluidstructs\hspace{.1pt}\discretionary{.}{%
}{.}\hspace{.4pt}2013\hspace{.1pt}\discretionary{.}{%
}{.}\hspace{.4pt}08\hspace{.1pt}\discretionary{.}{%
}{.}\hspace{.4pt}007}}


\bibitem{Thompson2011}
D.~Thompson, J.~A. Levine, J.~C. Bennett, P.~T. Bremer, A.~Gyulassy,
  V.~Pascucci, and P.~P. P\'{e}bay.
\newblock Analysis of large-scale scalar data using hixels.
\newblock In {\em IEEE Symposium on Large Data Analysis and Visualization}, pp.
  23--30, 2011. doi: {{%
10\hspace{.1pt}\discretionary{.}{%
}{.}\hspace{.4pt}1109\discretionary{/}{%
}{/}LDAV\hspace{.1pt}\discretionary{.}{%
}{.}\hspace{.4pt}2011\hspace{.1pt}\discretionary{.}{%
}{.}\hspace{.4pt}6092313}}


\bibitem{Ulichney1988}
R.~A. {Ulichney}.
\newblock Dithering with blue noise.
\newblock {\em Proceedings of the IEEE}, 76(1):56--79, 1988. doi: {{%
10\hspace{.1pt}\discretionary{.}{%
}{.}\hspace{.4pt}1109\discretionary{/}{%
}{/}5\hspace{.1pt}\discretionary{.}{%
}{.}\hspace{.4pt}3288}}


\bibitem{Ulichney1993}
R.~A. Ulichney.
\newblock Void-and-cluster method for dither array generation.
\newblock In {\em Human Vision, Visual Processing, and Digital Display IV},
  vol. 1913, pp. 332--343, 1993. doi: {{%
10\hspace{.1pt}\discretionary{.}{%
}{.}\hspace{.4pt}1117\discretionary{/}{%
}{/}12\hspace{.1pt}\discretionary{.}{%
}{.}\hspace{.4pt}152707}}


\bibitem{Wang2017}
K.~C. Wang, K.~Lu, T.~H. Wei, N.~Shareef, and H.~W. Shen.
\newblock Statistical visualization and analysis of large data using a
  value-based spatial distribution.
\newblock In {\em IEEE Pacific Visualization Symposium}, pp. 161--170, 2017.
  doi: {{%
10\hspace{.1pt}\discretionary{.}{%
}{.}\hspace{.4pt}1109\discretionary{/}{%
}{/}PACIFICVIS\hspace{.1pt}\discretionary{.}{%
}{.}\hspace{.4pt}2017\hspace{.1pt}\discretionary{.}{%
}{.}\hspace{.4pt}8031590}}


\bibitem{Wei2018}
T.~Wei, S.~Dutta, and H.~Shen.
\newblock Information guided data sampling and recovery using bitmap indexing.
\newblock In {\em IEEE Pacific Visualization Symposium}, pp. 56--65, 2018. doi:
  {{%
10\hspace{.1pt}\discretionary{.}{%
}{.}\hspace{.4pt}1109\discretionary{/}{%
}{/}PacificVis\hspace{.1pt}\discretionary{.}{%
}{.}\hspace{.4pt}2018\hspace{.1pt}\discretionary{.}{%
}{.}\hspace{.4pt}00016}}


\bibitem{Woodring2011}
J.~Woodring, J.~Ahrens, J.~Figg, J.~Wendelberger, S.~Habib, and K.~Heitmann.
\newblock In-situ sampling of a large-scale particle simulation for interactive
  visualization and analysis.
\newblock {\em Computer Graphics Forum}, 30(3):1151--1160, 2011. doi: {{%
10\hspace{.1pt}\discretionary{.}{%
}{.}\hspace{.4pt}1111\discretionary{/}{%
}{/}j\hspace{.1pt}\discretionary{.}{%
}{.}\hspace{.4pt}1467\discretionary{%
}{-}{-}8659\hspace{.1pt}\discretionary{.}{%
}{.}\hspace{.4pt}2011\hspace{.1pt}\discretionary{.}{%
}{.}\hspace{.4pt}01964\hspace{.1pt}\discretionary{.}{%
}{.}\hspace{.4pt}x}}


\end{thebibliography}
\end{document}